\documentclass[5p,lefttitle]{elsarticle}

\usepackage{graphicx}  
\usepackage{booktabs}  
\usepackage{dsfont}    

\usepackage{listings}  
\title{Using the \texttt{bayesmeta} \textsf{R}~package
       for Bayesian random-effects meta-regression}

\author{Christian R\"{o}ver\corref{cor1}}
\ead{christian.roever@med.uni-goettingen.de}
\address{Department of Medical Statistics, University Medical Center G\"{o}ttingen, Humboldt\-allee~32, 37073~G\"{o}ttingen, Germany}

\author{Tim Friede}
\ead{tim.friede@med.uni-goettingen.de}
\address{Department of Medical Statistics, University Medical Center G\"{o}ttingen, Humboldt\-allee~32, 37073~G\"{o}ttingen, Germany}

\cortext[cor1]{Corresponding author}

\date{\texttt{+++ preprint version +++ November, 2022 +++}}

\providecommand{\normaldistn}{\mathrm{Normal}}

\providecommand{\differential}{\mathrm{d}}
\providecommand{\realline}{\mathds{R}}

\providecommand{\Sigmatau}{\Sigma_{\tau}}
\providecommand{\Vin}{$\bullet$} 
\providecommand{\Vout}{$\circ$}  

\begin{document}
  \begin{abstract}
    \textbf{Background:} Random-effects meta-analysis within a hierarchical normal modeling framework is commonly implemented in a wide range of evidence synthesis applications. More general problems may even be tackled when considering \emph{meta-regression} approaches that in addition allow for the inclusion of study-level covariables. \textbf{Methods:} We describe the Bayesian meta-regression implementation provided in the \texttt{bayesmeta} \textsf{R}~package including the choice of priors, and we illustrate its practical use. \textbf{Results:} A wide range of example applications are given, such as binary and continuous covariables, subgroup analysis, indirect comparisons, and model selection. Example \textsf{R}~code is provided.
    \textbf{Conclusions:} The \texttt{bayesmeta} package provides a flexible implementation. 
    Due to the avoidance of MCMC methods, computations are fast and reproducible, 
%
    facilitating quick sensitivity checks 
    or large-scale simulation studies.
  \end{abstract}
  \maketitle

  \setlength\emergencystretch{\hsize}
  \section{Introduction}
    In the course of scientific endeavour it is often necessary to assess the compiled evidence from several separate sources, e.g., from independent experiments. \emph{Meta-analysis} methods have emerged as a popular class of tools to perform such evidence syntheses, which are now\-adays commonplace in many scientific disciplines \citep{ChalmersHedgesCooper2002,GurevitchEtAl2018}.

    A simple, versatile and common approach to meta-analysis is given by the \emph{normal-normal hierarchical model (NNHM)}, where measurement uncertainty as well as variability between measurements are implemented using normal distributions \citep{HartungKnappSinha,HedgesOlkin}. Inference within the NNHM framework may be tackled in different ways, and a Bayesian approach has proven particularly useful \citep{Schmid2001,SmithSpiegelhalterThomas1995,SuttonAbrams2001,WeltonJonesDias2020}. The technical implementation is commonly facilitated using Mar\-kov chain Monte Carlo (MCMC) methods \citep{McmcInPractice}. However, the relatively simple NNHM also lends itself to a semi-analytical solution using the \textsc{direct} algorithm \citep{RoeverFriede2017}.  Meta-analysis within the generic NNHM is implemented in the \texttt{bayesmeta} \textsf{R}~package \citep{bayesmeta,Roever2020}.

    The simple NNHM is readily generalized to a \emph{meta-regression} model that allows for the consideration of additional covariables (at the level of individual estimates, i.e. the level of the studies or experiments) in a meta-analysis \citep{HigginsEtAl2021,LauIoannidisSchmid1998,ThompsonHiggins2002,TiptonEtAl2019a}. This common model extension may again also be analyzed via the \textsc{direct} approach, technically by extending from one-dimensional (conditional or marginal) posterior distributions of a single ``overall mean'' or ``intercept'' parameter to higher-dimensional posterior distributions of a set of regression coefficients.  This approach was recently implemented and included in the \texttt{bayesmeta} \textsf{R}~package; the present paper gives an overview of the new functionality and showcases its application in a range of different analysis scenarios.

    Meta-regression methods aim to attribute differences apparent between individual empirical estimates to available covariables, and with that will reduce the between-study variance (heterogeneity) --- just like consideration of additional covariables in an ordinary regression will generally improve the model fit and increase the \emph{coefficient of determination} \citep{Mahoney2004}. Meta-regression analyses are hence often seen in the context of the exploration of (potential) sources of heterogeneity \citep{Higgins2008,Thompson1994}, with the intention to reduce or eliminate any unexplained variance and reduce bias \citep{BakerEtAl2009,DiasEtAl2013c,MortonEtAl2004,Schmid1999}. However, \citet{ThompsonHiggins2002} caution that associations derived from meta-regression are \emph{observational} in nature and that data dredging may be an issue, while \citet{HartungKnappSinha} also point out the danger of overfitting due to the commonly small sample sizes (numbers of studies). 
    With that, the \emph{statistical power} to identify covariables will commonly also tend to be low \citep{HempelEtAl2013}.
    \citet{Cooper2009} in this context distinguishes between the concepts of \emph{study-generated} and \emph{synthesis-generated evidence}, and cautions that in general only the former may allow to infer causal relationships.
    Such issues might to some extent be addressed by pre-specification of analyses \citep{StewartMoherShekelle2012}, while caution should be advised in general in order to avoid methodological problems (such as \emph{ecological fallacy}) \citep{GeissbuehlerEtAl2021}.

    While the scope of meta-regression methods is very broad, in practice a large number of practical applications are concerned with the investigation of \emph{subgroups} of estimates, which on the technical side means the consideration of binary ``indicator'' covariables. Such situations are often dealt with by simply analyzing groups of studies separately or jointly \citep{DoneganEtAl2015,KontopantelisSpringateReeves2013}. Meta-regression provides an alternative approach by allowing for differences in subgroup means while assuming a common heterogeneity variance. The use of meta-regression methods here has the advantage that questions regarding \emph{differences} between subgroup means are readily addressed, and that in particular in case of ``small'' subgroups, pathological behaviour is avoided through the borrowing of information on the heterogeneity (nuisance) parameter that is effectively taking place \citep{DiasEtAl2013c,DoneganEtAl2015,CochraneHandbook,TiptonEtAl2019a}.
    
    The meta-regression implementation described here facilitates a range of applications, including parameter estimation, prediction, shrinkage estimation, indirect comparisons, sensitivity analyses, model selection or model averaging. In the following, familiarity with some basics of (Bayesian) random-effects meta-analysis is a bonus \citep{Roever2020}, but is not strictly necessary. Methods will be introduced with a focus on applications rather than on theoretical background. The remainder of the manuscript is structured as follows. In Section~\ref{sec:rema}, the random-effects meta-regression model (i.e. NNHM with covariables) is introduced, including more guidance on model specification details, in particular the 
    parametrization
     of covariable effects and prior distributions, while some technical details and aims are also covered. Example applications follow in Section~\ref{sec:exampleApplications};  here we provide a wide range of potential applications with code snippets illustrated by real data. We close with a brief discussion in Section~\ref{sec:discussion}.


   \section{Methods} \label{sec:rema}
    \subsection{The data model}
    Meta-regression can be facilitated through a generalization of the normal-normal hierarchical model (NNHM) that is commonly used for random-effects meta-analysis \citep{Roever2020,WeltonJonesDias2020}. The model here is extended in order to consider linear effects of a set of study-level covariables.

    Suppose that a set of estimates from $k$~studies are to be modelled.  We then have $k$~estimates~$y_i \in \realline$ (where $i=1,\ldots,k$) with standard errors~$\sigma_i \in \realline_+$, which are assumed known. For each of the $k$~estimates, we also have a set of corresponding covariables $x_i\in\realline^d$ of dimension~$d$.
    Such (study-level) covariables are sometimes also denoted as \emph{moderators}.

    It is then assumed that each estimate quantifies an underlying parameter~$\theta_i$ with a normally distributed offset whose magnitude depends on the standard error~$\sigma_i$:
    \begin{eqnarray}\label{eqn:model1}
      y_i | \theta_i, \sigma_i & \sim & \normaldistn(\theta_i, \, \sigma_i^2) \mbox{.}
    \end{eqnarray}
    The study-specific mean~($\theta_i$) then depends linearly on the covariables~$x_i$ via a $d$-dimensional coefficient vector~$\beta$. However, even for an identical set of covariables, the mean may vary from study to study due to additional (normally distributed) variability:
    \begin{eqnarray}\label{eqn:model2}
      \theta_i | x_i, \beta, \tau & \sim & \normaldistn(\beta_1x_{i1}+\ldots+\beta_dx_{id}, \, \tau^2) \mbox{.}
    \end{eqnarray}
    Besides measurement or sampling errors ($\sigma_i$), the between-study variation is hence determined both by effects of covariables~$x_i$ as well as by \emph{heterogeneity} that is quantified through~$\tau$.  The model may also be formulated via the marginal expression
    \begin{eqnarray}\label{eqn:model3}
      y_i | x_i, \beta, \tau, \sigma_i  & \sim & \normaldistn(\beta_1x_{i1}+\ldots + \beta_dx_{id}, \nonumber \\
      & & \hspace{9ex} \sigma_i^2 + \tau^2) \mbox{.}
    \end{eqnarray}

    It is often convenient to alternatively view the model in vector/matrix terminology; here we may re-write equations (\ref{eqn:model1})--(\ref{eqn:model3}) as
    \begin{eqnarray}
      y | \theta, \sigma & \sim & \normaldistn(\theta, \, \Sigma) \nonumber\\
                                && \quad \mbox{where } \Sigma=\mathrm{diag}(\sigma_1^2,\ldots,\sigma_k^2)\mbox{,}\\
      \theta | X, \beta, \tau & \sim & \normaldistn(X \beta, \, \tau^2 I)\mbox{,}\quad\mbox{and} \\
      y | X, \beta, \tau, \sigma  & \sim & \normaldistn(X \beta, \, \Sigmatau) \nonumber\\
                                           && \quad \mbox{where } \Sigmatau=\Sigma + \tau^2I\mbox{,}
    \end{eqnarray}
    where the data are given in terms of the vectors of estimates $y\in\realline^k$ and standard errors~$\sigma \in \realline_+^k$, and the set of covariables forms the regressor matrix $X\in\realline^{k \times d}$, with rows corresponding to studies, and columns corresponding to different variables.

    The unknowns in the model are the study-specific effects~$\theta_i \in \realline$, the heterogeneity~$\tau \in \realline_+$, and the vector of coefficients $\beta\in\realline^{d}$ of dimension~${d}$. Prior distributions need to be specified for~$\tau$ and~$\beta$. In order to include an ``intercept'' (overall mean) term in the regression, one may specify one of the covariables (e.g., the first column of~$X$) as $x_{i1} = 1$ for $i=1,\ldots,k$.  Also, if \emph{only} an intercept term is considered, the model again simplifies to the ``plain'' meta-analysis model \citep{Roever2020}.


  \subsection{Prior and data specification}\label{sec:modelSpecification}
    \subsubsection{Effect and heterogeneity priors}
    Prior specification works similarly to the simple random-effects meta-analysis model \citep{Roever2020}; guidances provided for sensible specifications of heterogeneity ($\tau$) priors largely apply here as well \citep{Roever2020,RoeverEtAl2021}.

    Due to the implementation used in the \texttt{bayesmeta} package, priors for the $\beta$~coefficients may be specified as (proper) multivariate normal or (improper) uniform only. These generic forms will however be appropriate to cover a majority of common applications.


    \subsubsection{Alternative model specifications}
    What becomes crucial in addition then is the \emph{model specification}, i.e., the setup of covariables~$x_i$ eventually constituting the regressor matrix (or \emph{design matrix})~$X$.  Different, and to some extent equivalent, conventions are conceivable in order to approach the same analysis problem. This holds in particular in the context of binary covariables, which then imply different interpretations of the associated parameters (coefficients) and with that sometimes differing prior settings. A simple example involving binary covariables is illustrated in Table~\ref{tab:regressorCoding}.
    \begin{table}[t]
      \caption{\label{tab:regressorCoding}Illustration of different popular binary regressor matrix~($X$) codings for an example setting involving $k=7$ studies in $d=3$ subgroups. On the left-hand side, the three regression parameters $\beta_1$ to $\beta_3$ simply correspond to the three groups' means. On the right-hand side, $\beta_1^\star$ again corresponds to the mean in group~A, which also serves as a ``reference''. Parameters $\beta_2^\star$ and $\beta_3^\star$ correspond to the differences (``contrasts'') between the means within groups~A and~B and groups~A and~C, respectively.}
      \centering
      \begin{tabular}{cccccccccc}
        \toprule
               &&& \multicolumn{3}{c}{\small \it ``group mean''} 
               && \multicolumn{3}{c}{\small \it ``intercept/offset''} \\
               &&& \multicolumn{3}{c}{\small \it parametrisation} 
               && \multicolumn{3}{c}{\small \it parametrisation} \\
        \cmidrule(lr){4-6}
        \cmidrule(lr){8-10}
        $i$ & subgroup & & $x_{i1}$ & $x_{i2}$ & $x_{i3}$ & & $x_{i1}^\star$ & $x_{i2}^\star$ & $x_{i3}^\star$\\
        \midrule
        $1$ & A && $1$ & $0$ & $0$ && $1$ & $0$ & $0$ \\
        $2$ & A && $1$ & $0$ & $0$ && $1$ & $0$ & $0$ \\
        $3$ & B && $0$ & $1$ & $0$ && $1$ & $1$ & $0$ \\
        $4$ & B && $0$ & $1$ & $0$ && $1$ & $1$ & $0$ \\
        $5$ & B && $0$ & $1$ & $0$ && $1$ & $1$ & $0$ \\
        $6$ & C && $0$ & $0$ & $1$ && $1$ & $0$ & $1$ \\
        $7$ & C && $0$ & $0$ & $1$ && $1$ & $0$ & $1$ \\
        \bottomrule
      \end{tabular}
    \end{table}
    Suppose three groups of studies (labelled A, B and~C) are given, which may be identified using indicator variables as regressors in the design matrix~$X$. Two possible setups are shown here; on the left-hand side, the first study's mean is modeled as (see equation~(\ref{eqn:model3})) $\beta_1x_{11}+\beta_2x_{12}+\beta_3x_{13} = \beta_1$, whereas the third study's mean is $\beta_1x_{31}+\beta_2x_{32}+\beta_3x_{33} = \beta_2$. The three $\beta$~parameters hence directly correspond to the three group means.
    On the right-hand side, the first study is modelled in the same way, however, the third study's mean is $\beta_1x_{31}+\beta_2x_{32}+\beta_3x_{33} = \beta_1 + \beta_2$. The second coefficient ($\beta_2$) hence corresponds to the \emph{difference} (or \emph{contrast}) between groups~A and~B\@, while group~A serves as a ``reference''.
    Either way of formulating models may have its merits, and switching from one to another corresponds to a transformation between differing parameter spaces \citep[Sec.~1.8]{BDA3rd}. Prior settings for the regression coefficients may have differing implications in different model set\-ups, and prior settings in one parametrisation may be translated to another by considering the transformation step. 

    More specifically, in the example from Table~\ref{tab:regressorCoding}, the right-hand side (``intercept / offset'') parametrization results from the left-hand side (``group means'') via a linear transformation $\beta \rightarrow \beta^\star$, where
    \begin{eqnarray}
      \beta^\star & = & 
      \left(\begin{array}{c}
              \beta_1^\star \\ \beta_2^\star \\ \beta_3^\star
            \end{array}\right)
      \;=\;
      \left(\begin{array}{c}
              \beta_1 \\ \beta_2-\beta_1 \\ \beta_3-\beta_1
            \end{array}\right)
      \nonumber\\
      &=&
      \left(\begin{array}{rrr}
              1 & 0 & 0 \\ -1 & 1 & 0 \\ -1 & 0 & 1
            \end{array}\right)
      \left(\begin{array}{c}
              \beta_1 \\ \beta_2 \\ \beta_3
            \end{array}\right) 
      \;=\; A\beta
      \mbox{.}
    \end{eqnarray}
    Implications of prior assumptions in one parametrization may be derived by considering the transformation's effect on the transformed random variable \citep[Sec.~1.8]{BDA3rd}.
    In this case, if a normal prior with mean~$\mu$ and variance~$\Sigma$ was assumed for~$\beta$, then this implies a normal prior with mean~$A\mu$ and variance~$A \Sigma A^\prime$ for the transformed parameter set~$\beta^\star$.

    Transformations between alternative parametrisations, which may also be the result of transformations of the original \emph{data}, are often useful to simplify interpretation, or in order to avoid numerical problems. These may also arise in the context of continuous covariables, for example, when re-expressing fractions as percentages, or when ``centering'' covariables by subtracting their mean levels.


  \subsection{Inference} \label{sec:inference}
    \subsubsection{Technical implementation}
      In the \texttt{bayesmeta} \textsf{R}~package, the \textsc{direct} algorithm is utilized to facilitate meta-analysis within the NNHM framework via the \texttt{bayesmeta()} function \citep{RoeverFriede2017}. In contrast to the ``simple'' meta-analysis setup considered previously by \citet{Roever2020}, instead of a single ``overall mean'' parameter~$\mu$, the meta-\emph{regression} model now involves a $d$\mbox{-}dimensional coefficient vector~$\beta$, which means that some analytic expressions need to be generalized to their multivariate analogues; the basic algorithm for deriving the posterior distributions however may still be applied analogously.

      The meta-regression functionality is provided by the new \texttt{bmr()}~function; its main input arguments are:
      \begin{description}
        \item[\texttt{y}:] a vector of estimates ($y_i$) of length~$k$
        \item[\texttt{sigma}:] a vector of associated standard errors ($\sigma_i$) of length~$k$
        \item[\texttt{X}:] a regressor matrix ($X$) 
        with $k$~rows and $d$~columns
        \item[\texttt{tau.prior}:] a prior density function~($f(\tau)$) for the heterogeneity~$\tau$ (\emph{or} a character string denoting a specific form)
        \item[\texttt{beta.prior.mean}:] a vector of prior means of dimension~$d$
        \item[\texttt{beta.prior.sd}:] a vector of prior standard deviations of dimension~$d$
      \end{description}
      To a large extent, the behaviour is similar to the \texttt{bayesmeta()} function, especially with respect to the \texttt{y}, \texttt{sigma} and \texttt{tau.prior} arguments \citep{Roever2020}. The major differences to the \texttt{bayesmeta()} function are that one may specify an additional ``\texttt{X}''~argument giving the regressor matrix~($X$), and that the posterior, instead of referring to only a single effect~$\mu$, now involves a $d$\mbox{-}di\-men\-sion\-al parameter vector~$\beta$. By default, if the \texttt{tau.prior}, \texttt{beta.prior.mean} and \texttt{beta.prior.sd} arguments are left unspecified, (improper) uniform priors are assumed for~$\tau$ and~$\beta$. If an \texttt{X}~argument is not supplied, a single-column matrix of~ones is used, so that the analysis simplifies to fitting a single ``intercept'' parameter.

      Inference is facilitated through a semi-analystical approach, which is based on noting that the problem essentially involves two parameters, namely, the heterogeneity~$\tau$ and the coefficient vector~$\beta$. For any fixed heterogeneity value, the \emph{conditional} posterior distribution~$p(\beta|\tau)$ results analytically as a (multivariate) normal distribution. The heterogeneity's \emph{marginal} posterior density function~$p(\tau)$ on the other hand may also be expressed in analytical form. Noting that the  joint posterior density may be written as a product, i.e.,  $p(\beta,\tau)=p(\beta|\tau)p(\tau)$, then implies that the coefficients' marginal posterior results as a (continuous, normal) mixture distribution, i.e., $p(\beta)=\int p(\beta,\tau) \differential \tau = \int p(\beta|\tau) p(\tau) \differential \tau$. The \textsc{direct} algorithm utilized here for posterior computations then approximates the continuous mixture distribution by a discrete mixture using a finite number of components; a strategic setup of the set of support points then allows to control the computational accuracy \citep{RoeverFriede2017}.


    \subsubsection{Aims}
      Inference within a meta-regression application may be aimed at a range of different aspects, e.g., joint or marginal distributions of regression coefficients~($\beta_i$), linear combinations of coefficients~($x^\prime \beta$), investigation of heterogeneity~($\tau$), shrinkage estimation~($\theta_i$), or prediction~($\theta_{k+1}|x$). Posterior distributions of all these figures are available from the \texttt{bmr()} function's output. It is possible to access these directly from the object returned by the \texttt{bmr()} function, however, many relevant figures are included in the default output, and it is often convenient to request certain figures to be included e.g. in a summary printout or a forest plot. Ways to retrieve such figures are illustrated alongside the example applications below.


  \section{Results}  \label{sec:exampleApplications}
  In this section a number of applications are presented to illustrate the versatile use of meta-regression with \texttt{bayesmeta}.
    \subsection{Binary covariable}\label{sec:exampleBinary}
      \subsubsection{Inferring two means}
        \citet{CrinsEtAl2014} reported on a meta-analysis of studies investigating the use of inter\-leu\-kin-2 receptor antagonists (IL-2RA) for immunosuppression in pediatric liver transplant recipients. Of primary interest was the occurrence of \emph{acute rejection (AR)} reactions, a common adverse event that is supposed to be prevented by the medication. Two different types of IL-2RAs were used, namely, \emph{basiliximab} and \emph{daclizumab}. The rates of AR~events in the studies' treatment and control groups are summarized in terms of odds ratios (see also~\citep{Roever2020}); the relevant data are shown in Table~\ref{tab:CrinsData}.

        \begin{table}[t!]
        \begin{center}
          \caption{\label{tab:CrinsData}Data from the immunosuppression example. Each row here summarizes a $2\!\times\!2$ contingency table in terms of a derived log-OR~($y_i$) and its associated standard error~($\sigma_i$). Two different types of IL-2RA treatments were investigated (\emph{basiliximab} and \emph{daclizumab}).}
          \vspace{1ex}
          \begin{tabular}{cllcc}
            \toprule
            \multicolumn{2}{c}{study} &  & \multicolumn{2}{c}{log-OR}\\
            \cmidrule(lr){1-2} \cmidrule(lr){4-5}
            $i$ & reference & IL-2RA & $y_i$ & $\sigma_i$ \\
            \midrule
            $1$ & Heffron (2003)  & daclizumab  & $-2.31$ & $0.60$ \\
            $2$ & Gibelli (2004)  & basiliximab & $-0.46$ & $0.56$ \\
            $3$ & Schuller (2005) & daclizumab  & $-2.30$ & $0.88$ \\
            $4$ & Ganschow (2005) & basiliximab & $-1.76$ & $0.46$ \\
            $5$ & Spada (2006)    & basiliximab & $-1.26$ & $0.64$ \\
            $6$ & Gras (2008)     & basiliximab & $-2.42$ & $1.53$ \\
            \bottomrule
          \end{tabular}
        \end{center}
        \end{table}
        
        We will perform a meta-analysis aiming to investigate the mean effects for the two types of treatment; to that end, we specify the regressor matrix~$X$ reflecting the grouping of the data:
        \begin{equation}\label{eqn:CrinsRegressorMatrix}
          X \;=\; {\footnotesize
          \left(\begin{array}{cc}
                  0 & 1 \\ 1 & 0 \\ 0 & 1 \\
                  1 & 0 \\ 1 & 0 \\ 1 & 0
                \end{array}\right)}
          \mbox{.}
        \end{equation}
        In analogy to the coding illustrated in Table~\ref{tab:regressorCoding}, rows correspond to the six observations, and columns correspond to the two groups; the placement of zeroes and ones reflects the studies' associations to one of the two medication types. The investigation of differences in treatment effects here technically implies the consideration of a possible \emph{interaction} between treatment and IL-2RA~type. For the heterogeneity parameter~($\tau$), a half-normal prior distribution is appropriate in the context of a log-OR endpoint \citep{FriedeRoeverWandelNeuenschwander2017a,Roever2020,RoeverEtAl2021}. For the regression coefficients~($\beta_1$ and $\beta_2$), an (improper) uniform prior is used. First, the package and the example data set need to be loaded, and the effect measures (log-ORs) may be derived using the \texttt{metafor} package's \texttt{escalc()} function:
\begin{lstlisting}[language=R]
R> library("bayesmeta")
R> data("CrinsEtAl2014")
R> crins.es <- escalc(measure="OR",
+    ai=exp.AR.events,  n1i=exp.total,
+    ci=cont.AR.events, n2i=cont.total,
+    slab=publication, data=CrinsEtAl2014)
\end{lstlisting}
        Then we may specify the regressor matrix~$X$:
\begin{lstlisting}[language=R]
R> basil <- (crins.es$IL2RA=="basiliximab")
R> dacli <- (crins.es$IL2RA=="daclizumab")
R> X <- cbind("basiliximab"=as.numeric(basil),
+             "daclizumab" =as.numeric(dacli))
R> X
     basiliximab daclizumab
[1,]           0          1
[2,]           1          0
[3,]           0          1
[4,]           1          0
[5,]           1          0
[6,]           1          0
\end{lstlisting}
        The \texttt{bmr()} function then works very similarly to the \texttt{bayesmeta()} function \citep{Roever2020}; estimates and standard errors may be specified via the ``\texttt{y}'' and ``\texttt{sigma}'' arguments, or the data may simply be supplied in terms of the object returned from the \texttt{escalc()} function.  The heterogeneity prior is specified in terms of its probability density function, and in addition the regressor matrix needs to be provided via the ``\texttt{X}'' argument. We may hence specify
\begin{lstlisting}[language=R]
R> bmr01 <- bmr(y=crins.es$yi, 
+    sigma=sqrt(crins.es$vi), X=X,
+    tau.prior=function(t){dhalfnormal(t,s=0.5)})
\end{lstlisting}
        or simply
\begin{lstlisting}[language=R]
R> bmr01 <- bmr(crins.es, X=X,
+    tau.prior=function(t){dhalfnormal(t,s=0.5)})
\end{lstlisting}
        We may then have a closer look at the analysis output:
\begin{lstlisting}[language=R]
R> bmr01
 'bmr' object.

6 estimates:
Heffron (2003), Gibelli (2004), Schuller (2005),
Ganschow (2005), Spada (2006), Gras (2008)

2 regression parameters:
basiliximab, daclizumab

tau prior (proper):
function(t){dhalfnormal(t, scale=0.5)}
<bytecode: 0x555bbca28478>

beta prior: (improper) uniform

MAP estimates:
                  tau basiliximab daclizumab
joint    2.183638e-05   -1.287690  -2.307431
marginal 0.000000e+00   -1.283294  -2.307270

marginal posterior summary:
                tau basiliximab daclizumab
mode      0.0000000  -1.2832939 -2.3072701
median    0.2975462  -1.2832867 -2.3072414
mean      0.3420306  -1.2837755 -2.3072091
sd        0.2485756   0.3827227  0.5842337
95% lower 0.0000000  -2.0407937 -3.4589967
95% upper 0.8130176  -0.5252345 -1.1554177

(quoted intervals are shortest credible intervals.)
\end{lstlisting}
        The function's output again is very similar to the \texttt{bayesmeta()} function's output (see also~\citep{Roever2020}); at the top we see details of the model specification, the number and labels for the included studies, the number of regression coefficients and the prior specification. The variable names (here: ``basiliximab'' and ``daclizumab'') were taken from the column names of the regressor matrix.  Then maximum-a-posteriori (MAP) estimates are shown, as well as summary statistics for the three parameters' marginal posterior distributions.  The median and 95\% CIs for the \emph{basiliximab} and \emph{daclizumab} parameters ($\beta_1$ and $\beta_2$) are given by $-1.38$ [$-2.04$, $-0.53$] and $-2.31$ [$-3.46$, $-1.16$], respectively. The treatment hence appears to be effective in reducing AR events in both study groups.

        The (here: three) parameters' marginal or joint posterior distributions may also be inspected using the \texttt{plot()} or \texttt{pairs()} functions. The posterior distributions may also be accessed e.g.\ via the functions contained in the returned object's elements; for example, the \texttt{bmr01\$qposterior()} function allows to compute posterior quantiles. A call of
\begin{lstlisting}[language=R]
R> bmr01$qposterior(tau.p=0.99)
[1] 1.072312
\end{lstlisting}
        returns the heterogeneity posterior's 99\% quantile. Similarly, using
\begin{lstlisting}[language=R]
R> bmr01$qposterior(beta.p=0.99, which.beta=1)
[1] -0.3548454
\end{lstlisting}
        the 99\% quantile of the $\beta_1$~parameter may be determined. The ``\texttt{which.beta}'' argument here is used to specify the $\beta$~parameter's index. Analogously, the ``\texttt{...\$dposterior()}'', ``\texttt{...\$pposterior()}'', ``\texttt{...\$rposterior()}'' and ``\texttt{...\$post.interval()}'' functions may be used to determine posterior density, cumulative distribution function, random numbers or credible intervals (the naming of functions here follows the common \textsf{R}~conventions, as e.g.\ known from the \texttt{dnorm()} or \texttt{pnorm()} functions).
 
        Estimates of the $\theta_i$ parameters ($i=1,\ldots,k$), the \emph{shrinkage estimates}, may also be derived. Some summary statistics are provided in the ``\texttt{...\$theta}'' element. Access to the complete distributions (probability density, cumulative distribution function, quantile function, random number generation and credible intervals) is provided via the ``\texttt{...\$dshrink()}'', ``\texttt{...\$pshrink()}'', ``\texttt{...\$qshrink()}'', ``\texttt{...\$rshrink()}'' and ``\texttt{...\$shrink.interval()}'' functions.

\begin{figure}[b!]
          \centering
          {\includegraphics[width=\linewidth]{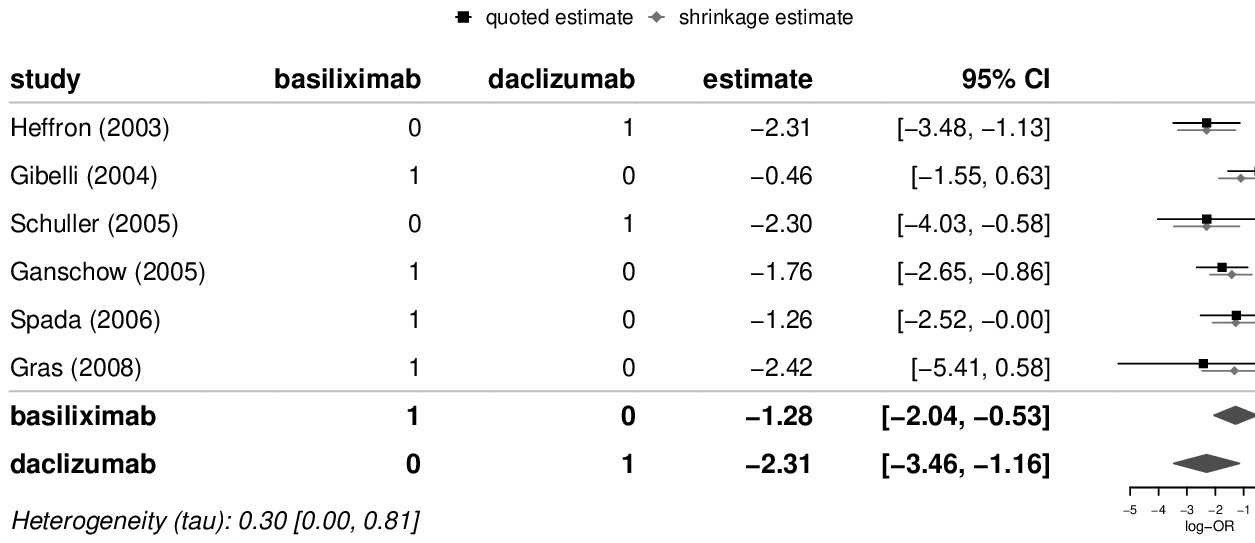}}
          \caption{\label{fig:CrinsForest01} A forest plot illustrating the meta-regression results based on the transplantation data from Table~\ref{tab:CrinsData}. The two study groups are coded in terms of two binary indicator variables labelled ``basiliximab'' and ``daclizumab''.}
        \end{figure}

        To illustrate the results in a forest plot, one may call 
\begin{lstlisting}[language=R]
R> forestplot(bmr01, xlab="log-OR")
\end{lstlisting}
        The resulting plot is shown in Figure~\ref{fig:CrinsForest01}.
        The forest plot's first six lines show the original data (estimates $y_i$ and 95\% CIs based on standard errors~$\sigma_i$ and the normal model), and the shrinkage estimates ($\theta_i$). The table also includes the regressor matrix~$X$ in the columns that are labelled as ``basiliximab'' and ``daclizumab'', as in the original specification of the ``\texttt{X}'' argument.
        The two lines at the bottom then show the estimates of the two associated regression parameters $\beta_1$ and $\beta_2$. The heterogeneity ($\tau$) distribution finally is also summarized at the bottom left.

        Note that while in a simple meta-analysis shrinkage estimates are ``shrunk'' towards the common overall mean, in a meta-regression shrinkage acts in the direction of the corresponding predicted value; in this case this means that individual studies' shrinkage estimates move towards the corresponding (basiliximab or daclizumab) group means.


      \subsubsection{Inferring means, contrasts, or predictions}\label{sec:exampleContrast}
        Quite commonly, it is also of interest to evaluate the posterior distribution of \emph{linear combinations} of the regression coefficients ($\beta_i$), or of \emph{predictions} corresponding to such combinations. For such purposes, the  ``\texttt{...\$dpredict()}'', ``\texttt{...\$ppredict()}'', ``\texttt{...\$qpredict()}'', ``\texttt{...\$rpredict()}'' and ``\texttt{...\$pred.interval()}'' functions are available.

        In the present example, it may be of interest to infer the \emph{difference} of the two group means, \mbox{$\beta_2-\beta_1$}. If its posterior includes zero, then the two medications may be equally effective; if zero is outside the plausible range, this indicates differing efficacies. The above difference is a \emph{linear combination} of the two coefficients, i.e., a sum of $\beta_2-\beta_1\;=\,-1\!\times\!\beta_1 + 1\!\times\!\beta_2$ with coefficients $-1$ and $+1$ for $\beta_1$ and $\beta_2$, respectively. We can request the linear combination's distribution by specifying the two coefficients, e.g., in order to determine the median or a 95\% CI:
\begin{lstlisting}[language=R]
R> bmr01$qpredict(p=0.5, x=c(-1,1))
[1] -1.023955
R> bmr01$pred.interval(level=0.95, x=c(-1,1))
    lower     upper 
-2.402393  0.354489 
attr(,"interval.type")
[1] "shortest"
\end{lstlisting}
        Such coefficients were in fact already quoted along with the summary estimates in the forest plot in Figure~\ref{fig:CrinsForest01}, although only involving zeroes and ones as coefficients.  Analogously, additional linear combinations can be specified for the plot; e.g., to include the group difference in the figure, we may specify
\begin{lstlisting}[language=R]
R> forestplot(bmr01, xlab="log-OR",
+    X.mean=rbind("basiliximab"      = c( 1, 0),
+                 "daclizumab"       = c( 0, 1),
+                 "group difference" = c(-1, 1)))
\end{lstlisting}
        The resulting plot is shown in Figure~\ref{fig:CrinsForest02} (top).  It should be noted that the comparison of basiliximab vs. daclizumab constitutes an \emph{indirect comparison} here, as it contrasts two treatments that have not been ``directly'' compared in a head-to-head comparison in any of the six trials considered \citep{KieferEtAl2015}. For more on indirect comparisons, see also Section~\ref{sec:exampleNetworkMA} below.

        Besides the \emph{mean} effects, \emph{predictions} are often of interest, e.g., in order to assess plausible ranges for a ``future'' study's mean parameter~$\theta_{k+1}$ (which, in addition to the $\beta$~coefficients, also depends on the estimated amount of heterogeneity~$\tau$). In the present example, we might be interested in predicting the mean in a new study investigating basiliximab; we can check out some quantiles of the effect's distribution via
\begin{lstlisting}[language=R]
R> bmr01$qpredict(p=c(0.025, 0.5, 0.975), 
+                 x=c(1,0), mean=FALSE)
[1] -2.4611428 -1.2835800 -0.1125025
\end{lstlisting}
        We again need to specify the coefficients via the ``\texttt{x}'' argument, and the ``\texttt{mean}'' argument (which by default is \texttt{TRUE}) needs to be set to \texttt{FALSE} explicitly.


      \subsubsection{Alternative regressor matrix setups}
\begin{figure}[t]
          \centering{\includegraphics[width=\linewidth]{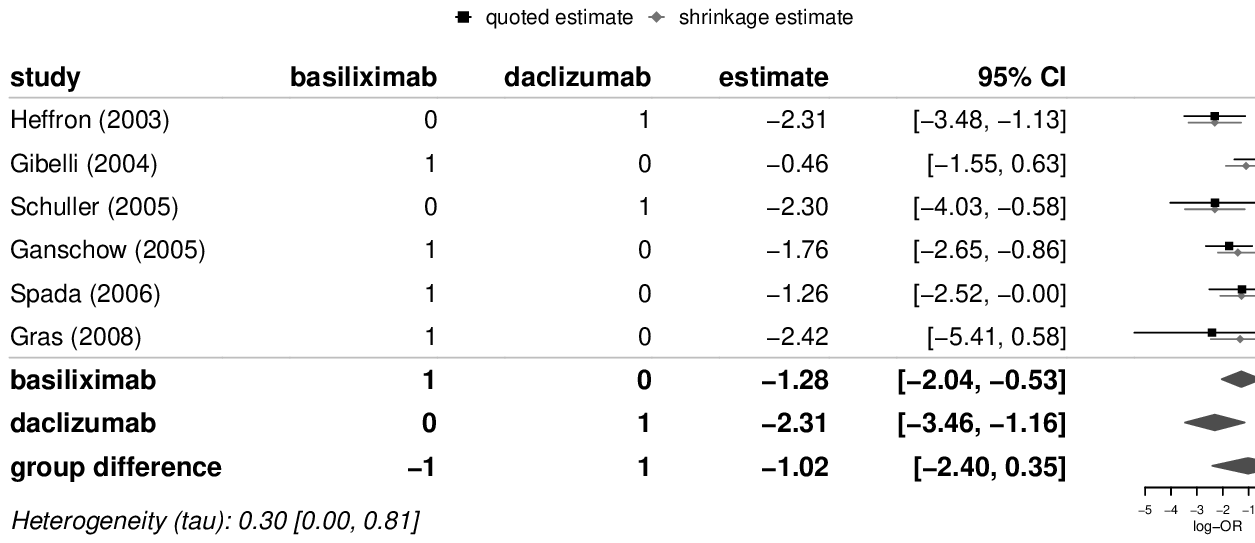}
                     \includegraphics[width=\linewidth]{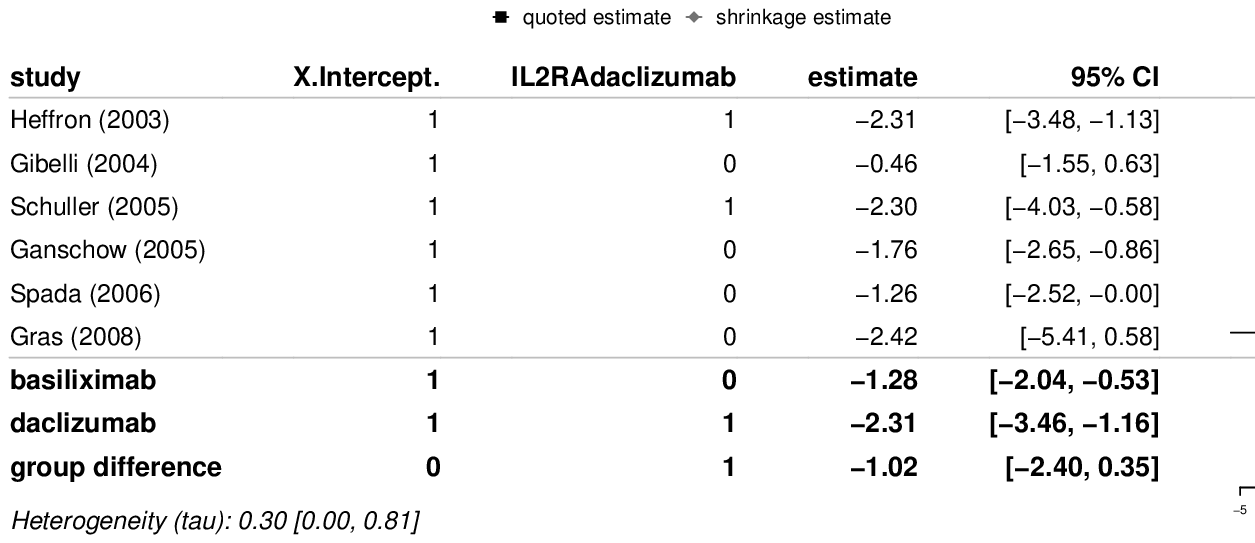}}
          \caption{\label{fig:CrinsForest02} Two forest plots similar to the one shown in Figure~\ref{fig:CrinsForest01} and illustrating analogous meta-regression results based on different parametrisations of the regressor matrix~$X$ (top: ``group mean'' parametrization, bottom: ``inter\-cept/slope'' parametrization; note the differing setups in the 2nd and 3rd ``regressor'' columns).}
        \end{figure}

        The specification of the regressor matrix~$X$ (see equation (\ref{eqn:CrinsRegressorMatrix}) above) is not unique; a number of different approaches are conceivable and common, for example, one might as well specify
        \begin{equation}
          X \;=\; {\footnotesize
          \left(\begin{array}{cc}
                  1 & 1 \\ 1 & 0 \\ 1 & 1 \\
                  1 & 0 \\ 1 & 0 \\ 1 & 0
                \end{array}\right)}
          \quad \mbox{or} \quad
          X \;=\; {\footnotesize
          \left(\begin{array}{cc}
                  1 & +0.5 \\ 1 & -0.5 \\ 1 & +0.5 \\
                  1 & -0.5 \\ 1 & -0.5 \\ 1 & -0.5
                \end{array}\right)}
        \end{equation}
        to yield analogous results. Different setups will then imply different interpretations for the associated $\beta$~parameters. Within~\textsf{R}, the most common parametrization is also returned by the \texttt{model.matrix()} function (from the \texttt{stats} package); if we run
\begin{lstlisting}[language=R]
R> X.alt <- model.matrix( ~ IL2RA, data=crins.es)
R> X.alt
  (Intercept) IL2RAdaclizumab
1           1               1
2           1               0
3           1               1
4           1               0
5           1               0
6           1               0
attr(,"assign")
[1] 0 1
attr(,"contrasts")
attr(,"contrasts")$IL2RA
[1] "contr.treatment"
\end{lstlisting}
        we can see that we in fact yield the first of the above two versions, an ``inter\-cept/slope'' parametrization, which would often be the default for many regression models within~\textsf{R}\@.  We may run the same analysis using this alternative regressor matrix:
\begin{lstlisting}[language=R]
R> bmr02 <- bmr(crins.es, X=X.alt,
+    tau.prior=function(t){dhalfnormal(t,s=0.5)})
\end{lstlisting}
        and generate a corresponding forest plot:
\begin{lstlisting}[language=R]
R> forestplot(bmr02, xlab="log-OR",
+    X.mean=rbind("basiliximab"      = c(1, 0),
+                 "daclizumab"       = c(1, 1),
+                 "group difference" = c(0, 1)))
\end{lstlisting}
        The resulting plot is shown in Figure~\ref{fig:CrinsForest02} (bottom).
        We can see that we get essentially identical results here, and that we only need to specify the linear combinations differently in order to retrieve group means or group differences based on the differing parametrizations.

        While in the above example the results are essentially equivalent (see also Section~\ref{sec:modelSpecification}), either way of formulating the problem may have its advantages. Interpretation of parameters and prior specification may be easier or more obvious in one or another formulation. In case informative priors for the regression parameters~$\beta$ were to be used in the above example, this may either imply considerations of constraints on the two individual group means, or on their difference.

        For example, consider the two alternative parametrizations in terms of
        \begin{equation}
          X \;=\; {\footnotesize
          \left(\begin{array}{cc}
                  1 & +0.5 \\ 1 & -0.5 \\ 1 & +0.5 \\
                  1 & -0.5 \\ 1 & -0.5 \\ 1 & -0.5
                \end{array}\right)}
          \quad \mbox{and} \quad
          X^\star \;=\; {\footnotesize
          \left(\begin{array}{cc}
                  0 & 1 \\ 1 & 0 \\ 0 & 1 \\
                  1 & 0 \\ 1 & 0 \\ 1 & 0
                \end{array}\right)}\mbox{.}
        \end{equation}
        The implied parameters may be thought of as ``overall mean / group difference'' or ``two group mean'' parameters.  The two associated parameter vectors~$\beta$ and~$\beta^\star$ are related to one another as
        \begin{eqnarray}
          \beta^\star &=&
          \left(\begin{array}{c}
                  \beta_1^\star \\ \beta_2^\star
                \end{array}\right)
          \;=\;
          \left(\begin{array}{c}
                  \beta_1 - 0.5\beta_2 \\ \beta_1+0.5\beta_2
                \end{array}\right)
          \;=\; A\beta\mbox{,} \nonumber\\
          && \mbox{with}\quad
          A\;=\;
          \left(\begin{array}{rrr}
                  1 & -0.5 \\ 1 & +0.5
                \end{array}\right)
          \mbox{.}
        \end{eqnarray}
        Now suppose that in the former parametrization ($\beta$) we want to implement a vague prior for the overall mean, while the difference between groups is expected to be rather small; for~$\beta$ we hence assume a normal prior with mean and covariance
        \begin{equation}
          \mu \;=\; {
          \left(\begin{array}{c}
                  0  \\ 0 
                \end{array}\right)}
          \qquad \mbox{and} \qquad
          \Sigma \;=\; {
          \left(\begin{array}{rr}
                  100 & 0 \\ 
                    0 & 1 
                \end{array}\right)}\mbox{.}
        \end{equation}
        The prior specification in the ``overall mean / group difference'' parametrization has its counterpart in the ``two group mean'' parametrization; for the ``transformed'' parameter~$\beta^\star$, this implies a prior distribution with mean and covariance
        \begin{eqnarray}
          \mu^\star &=& A\mu \;=\; {
          \left(\begin{array}{c}
                  0  \\ 0 
                \end{array}\right)}
          \qquad \mbox{and}\\
          \Sigma^\star &=& A\Sigma A^\prime \;=\; {
          \left(\begin{array}{rr}
                  100.25 &  99.75 \\ 
                   99.75 & 100.25
                \end{array}\right)}\mbox{.}
        \end{eqnarray}
        The high correlation reflects the assumption implemented in the original para\-metrization that there is a constraint on the difference between the means while their common average level has greater uncertainty.

        Performing the analysis using different parametrizations and matching proper, informative priors should then again yield identical sets of estimates as in the previous example (Figure~\ref{fig:CrinsForest02}). The two analyses may be performed via
\begin{lstlisting}[language=R]
R> bmr03 <- bmr(crins.es,
+    X=cbind("mean"=rep(1,6),
+            "difference"=c(0.5, -0.5, 0.5,
+                          -0.5, -0.5, -0.5)),
+    beta.prior.mean=c(0,0),
+    beta.prior.sd=c(10,1),
+    tau.prior=function(t){dhalfnormal(t,s=0.5)})
R> bmr04 <- bmr(crins.es, X=X,
+    beta.prior.mean=c(0,0),
+    beta.prior.cov=cbind(c(100.25, 99.75), 
+                         c(99.75, 100.25)),
+    tau.prior=function(t){dhalfnormal(t,s=0.5)})
\end{lstlisting}
        We may then check the corresponding estimates (of the two group means, the overall mean and the group difference) via the \texttt{summary()} function, which allows to specify an ``\texttt{X.mean}'' argument, similarly to the \texttt{forestplot()} function; for example:
\begin{lstlisting}[language=R]
R> smry03 <- summary(bmr03, 
+    X.mean=rbind("basiliximab" = c(1, -0.5),
+                 "daclizumab"  = c(1, +0.5),
+                 "average"     = c(1, 0),
+                 "difference"  = c(0, 1)))
R> smry04 <- summary(bmr04, 
+    X.mean=rbind("basiliximab" = c(1, 0),
+                 "daclizumab"  = c(0, 1),
+                 "average"     = c(0.5, 0.5),
+                 "difference"  = c(-1, 1)))
\end{lstlisting}
        We may first double check the differing setups of the underlying regressor matrices:
\begin{lstlisting}[language=R]
R> smry03$X.mean
            mean difference
basiliximab    1       -0.5
daclizumab     1        0.5
average        1        0.0
difference     0        1.0
R> smry04$X.mean
            basiliximab daclizumab
basiliximab         1.0        0.0
daclizumab          0.0        1.0
average             0.5        0.5
difference         -1.0        1.0
\end{lstlisting}
        and then check the resulting estimates:
\begin{lstlisting}[language=R]
R> smry03$mean[,c(2,5,6)]
                median 95% lower  95% upper
basiliximab -1.3736003 -2.094845 -0.6670075
daclizumab  -2.0821175 -3.077314 -1.0710786
average     -1.7288302 -2.394382 -1.0591354
difference  -0.7062405 -1.811174  0.4232024
R> smry04$mean[,c(2,5,6)]
                median 95% lower  95% upper
basiliximab -1.3736003 -2.094845 -0.6670075
daclizumab  -2.0821175 -3.077314 -1.0710786
average     -1.7288302 -2.394382 -1.0591354
difference  -0.7062405 -1.811174  0.4232024
\end{lstlisting}
        and indeed the corresponding estimates are identical under both model variations (and different from those shown in Figure~\ref{fig:CrinsForest02}, which were based on uniform priors).


      \subsubsection{Connection to meta-analysis without covariables}
        Meta-regression is a generalization of a ``simple'' meta-analysis, and the simple meta-analysis again constitutes the special case of a regression that does not consider additional covariables (besides an overall ``intercept'' term). We may also use the \texttt{bmr()} function without covariables by simply omitting the ``\texttt{X}'' argument:
\begin{lstlisting}[language=R]
R> bmr05 <- bmr(crins.es,
+   tau.prior=function(t){dhalfnormal(t, s=0.5)})
\end{lstlisting}
        We can check the regressor matrix used internally (which here may be accessed as ``\texttt{bmr05\$X}'') to see that indeed this is by default a matrix consisting of a single column of ones, so that a single ``intercept'' parameter~$\beta_1$ is fitted.

        We may then compare results against those from the \texttt{bayesmeta()} function; comparing the estimates for the overall mean, we get
\begin{lstlisting}[language=R]
R> bma <- bayesmeta(crins.es,
+    tau.prior=function(t){dhalfnormal(t,s=0.5)})
R> cbind("bmr"      =bmr05$summary[,"intercept"],
+        "bayesmeta"=bma$summary[,"mu"])
                 bmr  bayesmeta
mode      -1.5771362 -1.5779534
median    -1.5830833 -1.5834970
mean      -1.5871400 -1.5873257
sd         0.3310698  0.3314634
95% lower -2.2465774 -2.2464090
95% upper -0.9341840 -0.9338935
\end{lstlisting}
        Differences between the two results are due to differences in numerical accuracy; if we check the number of support points used internally for the approximation of the posterior (via ``\texttt{str(bmr05\$support)}'' or ``\texttt{str(bma\$support)}''), we can see that \texttt{bmr()} uses 9~support points, while \texttt{bayesmeta()} yields a grid of 17~points.  The slight discrepancy arises since within the \texttt{bayesmeta()} function, the grid setup is determined based on the marginal distribution of the overall mean ($\mu$) as well as the shrinkage estimates ($\theta_i$) \citep{Roever2020}, while in \texttt{bmr()} only the regression coefficients' (multivariate) distribution is considered. If desired, accuracy may always be increased by adjusting the ``\texttt{delta}'' or ``\texttt{epsilon}'' parameters \citep{RoeverFriede2017}.

        In the two-group comparison discussed above (see e.g.\ Figure~\ref{fig:CrinsForest02}), the two group means (basiliximab and daclizumab) are estimated ``independently'' in some sense, i.e., the estimates from one group of studies only help estimating the other group's mean insofar as they provide information on the heterogeneity, but not on the actual \emph{location}. The difference to performing two completely separate analyses of both group means is the assumption of a common heterogeneity parameter for both groups. This provides another connection to the ``simple'' meta-analysis model (without additional covariables): the two group mean estimates may also be recovered by performing two separate meta-analyses and propagating only the heterogeneity information. Consider the estimate of the \emph{daclizumab} group, which was given by
\begin{lstlisting}[language=R]
R> bmr01$summary[, "daclizumab", drop=FALSE]
          daclizumab
mode      -2.3072701
median    -2.3072414
mean      -2.3072091
sd         0.5842337
95% lower -3.4589967
95% upper -1.1554177
\end{lstlisting}
        The same estimate may be derived by first performing the analysis for the \emph{basi\-lixi\-mab} group only, and then using the resulting heterogeneity posterior as the prior for the subsequent \emph{daclizumab} analysis:
\begin{lstlisting}[language=R]
R> basil <- (crins.es$IL2RA=="basiliximab")
R> dacli <- (crins.es$IL2RA=="daclizumab"))
R> bma.bas <- bayesmeta(crins.es[basil,],
+    tau.prior=function(t){dhalfnormal(t,s=0.5)})
R> bma.dac <- bayesmeta(crins.es[dacli,],
+    tau.prior=function(t){bma.bas$dposterior(
+                                        tau=t)})
R> bma.dac$summary[, "mu", drop=FALSE]
                  mu
mode      -2.3072954
median    -2.3072387
mean      -2.3072071
sd         0.5847048
95% lower -3.4596942
95% upper -1.1547187
\end{lstlisting}
        One can see that the results are essentially identical, with slight discrepancies that may be attributed to numerical differences.


    \subsection{Indirect comparisons in a treatment network}\label{sec:exampleNetworkMA}
      It became already evident in the previous example (Section~\ref{sec:exampleContrast}) that fitting individual coefficients for certain pairwise comparisons, along with the option to infer linear combinations of coefficients, allows to estimate certain \emph{indirect comparisons} \citep{KieferEtAl2015}. In fact, applicability of the meta-regression model to some degree extends into the domain of \emph{network meta-analysis (NMA)} \citep[Sec.~11.4.2]{CochraneHandbook}. However, the scope here is somewhat limited, insofar as the model is \emph{contrast-based}, only two-armed trials may be considered, and a single common heterogeneity parameter is assumed \citep{DiasAdes2016,SalantiEtAl2008,WhiteEtAl2019}.

      For illustration, we will consider the example data set due to \citet{BucherEtAl1997}, which includes studies providing evidence for both the \emph{direct} as well as the \emph{indirect} comparison of two treatments.
      \begin{figure}[b]
        \centering
        {\includegraphics[width=0.6\linewidth]{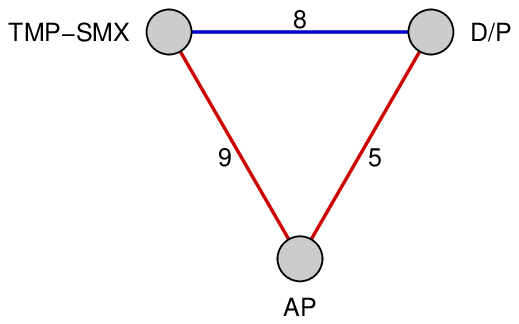}}
        \caption{\label{fig:BucherNetwork} Illustration of the triangular network of comparisons within the \citet{BucherEtAl1997} example data set.  8~studies provide a \emph{direct} head-to-head comparison of TMP-SMX vs.\ D/P (indicated by the blue edge); the remaining 14~studies (shown in red) provide \emph{indirect} evidence on the effect via the comparison of either TMP-SMX versus AP (9~studies), or D/P versus AP (5~studies).}
      \end{figure}
      \citeauthor{BucherEtAl1997} considered the example of the comparison of \emph{sulpha\-metoxa\-zole-tri\-metho\-prim (TMP-SMX)} versus \emph{dap\-sone/pyri\-metha\-mine (D/P)} for the prophylaxis of \emph{Pneumo\-cystis carinii} pneumonia (Pcp) in HIV patients. Eight studies had undertaken a head-to-head comparison of both medications, but an additional 14~studies were available investigating one of the medications with \emph{aero\-solized penta\-midine (AP)} as a comparator. Nine studies compared TMP-SMX vs.\ AP, and five studies compared D/P vs.\ AP\@. Together these provide \emph{indirect} evidence on the effect of TMP-SMX compared to D/P\@.  The resulting triangular network of pairwise comparisons is illustrated in Figure~\ref{fig:BucherNetwork}.

\begin{figure}[t]
        \centering
        {\includegraphics[width=\linewidth]{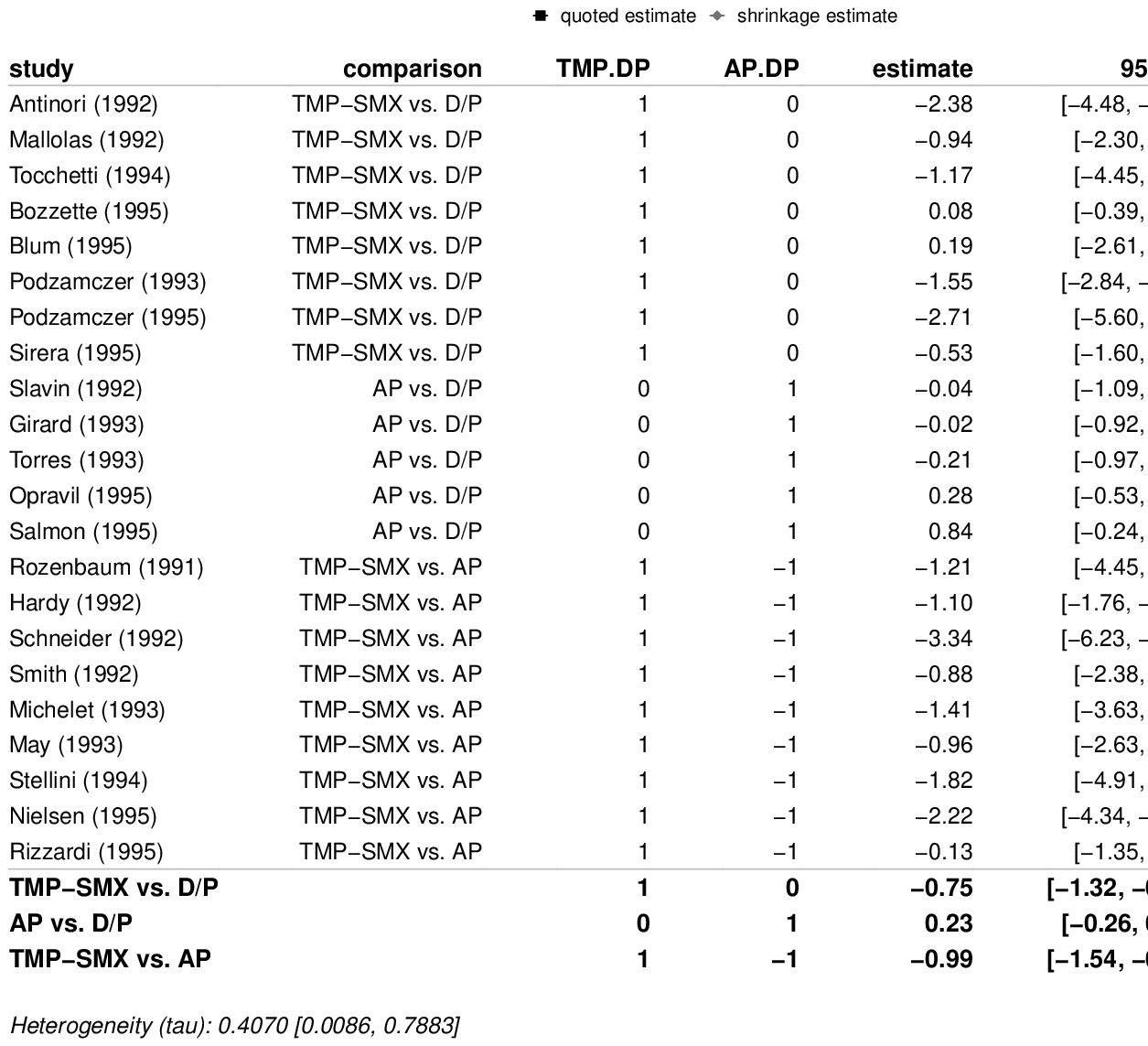}}
        \caption{\label{fig:BucherForest} Forest plot for the network-MA data set \citep{BucherEtAl1997}. The first 8~studies did a \emph{direct} head-to-head comparison of TMP-SMX vs.\ D/P\@; the remaining studies provide \emph{indirect} evidence on the effect via the comparison with AP\@. At the bottom, the estimates for all three pairwise comparisons are shown.}
      \end{figure}

      We may load the data and compute effect sizes (log-ORs) for all 22~studies.
\begin{lstlisting}[language=R]
R> data("BucherEtAl1997")
R> es <- escalc(measure="OR",
+               ai=events.A, n1i=total.A,
+               ci=events.B, n2i=total.B,
+               slab=study, data=BucherEtAl1997)
\end{lstlisting}
      We may then set up the regressor matrix to estimate the two relevant (non-redundant) treatment effects. Two coefficients ($\beta_1$ and $\beta_2$) are estimated; the first corresponds to the comparison of TMP-SMX against D/P, the second is for the comparison of AP again D/P, and the last remaining pairwise comparison (TMP-SMX vs.\ AP) then results as the difference of the former two.
\begin{lstlisting}[language=R]
R> X <- cbind("TMP.DP"=rep(c(1,0, 1), c(8,5,9)),
+             "AP.DP" =rep(c(0,1,-1), c(8,5,9)))
\end{lstlisting}
      Again, there are alternative (equivalent) ways to set up the regressor matrix \citep{SalantiEtAl2008}. The actual analysis then is performed via a call of
\begin{lstlisting}[language=R]
R> bmr06 <- bmr(es, X=X)
\end{lstlisting}
      The ``\texttt{tau.prior}'' argument is left unspecified, which means that the default of an (improper) uniform prior is used for~$\tau$, which should be appropriate given the reasonably large number of studies included here ($k=22$) \citep{RoeverEtAl2021}. Figure~\ref{fig:BucherForest} illustrates the data along with the regressor matrix and the estimated coefficients in a corresponding forest plot.

\begin{figure}[h]
        \centering
        {\includegraphics[width=0.9\linewidth]{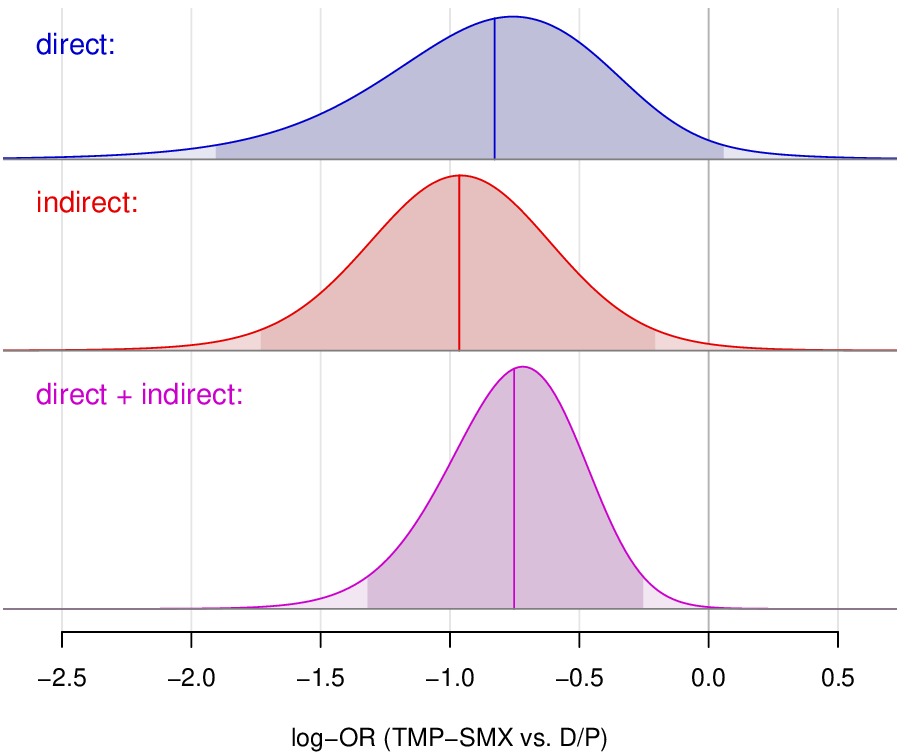}}
        \caption{\label{fig:BucherEstimates} Posterior distributions for the $\beta$~coefficient corresponding to the \mbox{log-OR} in the comparison of \emph{TMP-SMX} vs.\ \emph{D/P}, either considering only the studies providing direct or indirect evidence, or including all studies. Besides posterior densities, posterior medians and 95\% CIs are indicated.}
      \end{figure}

      In order to more closely investigate and contrast the ``direct'' and ``indirect'' contributions to specific estimates, we may have a closer look at the estimates resulting from considering subsets of the data. In addition to the analysis described above, we may run analyses based only on the subsets of studies providing direct or indirect evidence (studies 1--8 or studies 9--22):
\begin{lstlisting}[language=R]
R> bmr06.direct   <- bmr(es[1:8,], X=X[1:8,1])
R> bmr06.indirect <- bmr(es[9:22,], X=X[9:22,])
\end{lstlisting}
      In the overall analysis, the log-OR for the effect of TMP-SMX vs.\ D/P is estimated at $-0.75$ with 95\% CI [$-1.32$, $-0.25$]; the ``direct'' and ``indirect'' estimates are roughly similar and overlapping at $-0.83$ [$-1.91$, $0.06$] and $-0.96$ [$-1.73$, $-0.21$]. We may also inspect the corresponding posterior densities, which again are accesssible via the returned ``\texttt{\ldots\$dposterior()}'' functions.  In Figure~\ref{fig:BucherEstimates}, the three posteriors are contrasted side-by-side.
      All three estimates are consistent, and when combining direct and indirect evidence, the gain in precision becomes apparent.


    \subsection{Continuous covariable}
      \citet{NicholasEtAl2019} performed a systematic review and meta-analysis in order to examine how the characteristics of placebo groups of randomized controlled trials in multiple scelerosis may have evolved over time. A number of features were investigated, among these was the \emph{proportion of patients experiencing disability progression} within 24~months. 28~studies with available information on disability progression were found, spanning the period from the year~1990 until 2018\@. A time trend in the progression rates would mean a tendency towards more severe or more benign cases being investigated over the years, and may have implications for the comparability of results from older or more recent studies, or also for the design of future studies. We can load the example data, and from the studies' placebo group sizes and the observed percentages of progressing patients, we can compute estimates of the logarithmic odds of disease progression using the \texttt{escalc()} function:
\begin{lstlisting}[language=R]
R> data("NicholasEtAl2019")
R> head(NicholasEtAl2019)
              study year patients prog.percent
1 Kastrukoff (1990) 1990       50         46.0
2   Wolinsky (1990) 1990      274         72.5
3  Bornstein (1991) 1991       55         29.5
4    Likosky (1991) 1991       20         80.0
5 Noseworthy (1991) 1991       56         22.0
6   Milanese (1992) 1992       21         55.0
R> es <- escalc(measure="PLO",
+    xi=patients*(prog.percent/100), ni=patients,
+    slab=study, data=NicholasEtAl2019)
R> head(es[,c(1,2,5,6)])
              study year      yi     vi 
1 Kastrukoff (1990) 1990 -0.1603 0.0805 
2   Wolinsky (1990) 1990  0.9694 0.0183 
3  Bornstein (1991) 1991 -0.8712 0.0874 
4    Likosky (1991) 1991  1.3863 0.3125 
5 Noseworthy (1991) 1991 -1.2657 0.1041 
6   Milanese (1992) 1992  0.2007 0.1924 
\end{lstlisting}
      The ``\texttt{yi}'' and ``\texttt{vi}'' columns here give the log-odds and their (squared) standard errors. The (continuous) covariable of interest is given in the ``\texttt{year}'' column. We may then specify the regressor matrix:
\begin{lstlisting}[language=R]
R> X <- cbind("intercept2000" = 1,
+             "year" = (es$year-2000))
R> head(X)
     intercept2000 year
[1,]             1  -10
[2,]             1  -10
[3,]             1   -9
[4,]             1   -9
[5,]             1   -9
[6,]             1   -8
\end{lstlisting}
      Note that here we are using to a simple ``intercept/slope'' model, and that the ``year'' variable is re-coded so that the data are centered at the year~2000 (and the intercept parameter hence corresponds to the log-odds in~2000). We may then perform the analysis:
\begin{lstlisting}[language=R]
R> bmr07 <- bmr(es, X=X)
\end{lstlisting}
      A prior for the heterogeneity ($\tau$) again is not specified, implying that he default of an (improper) uniform prior is used \citep{RoeverEtAl2021}.

\begin{figure}[t!]
        \centering
        {\includegraphics[width=\linewidth]{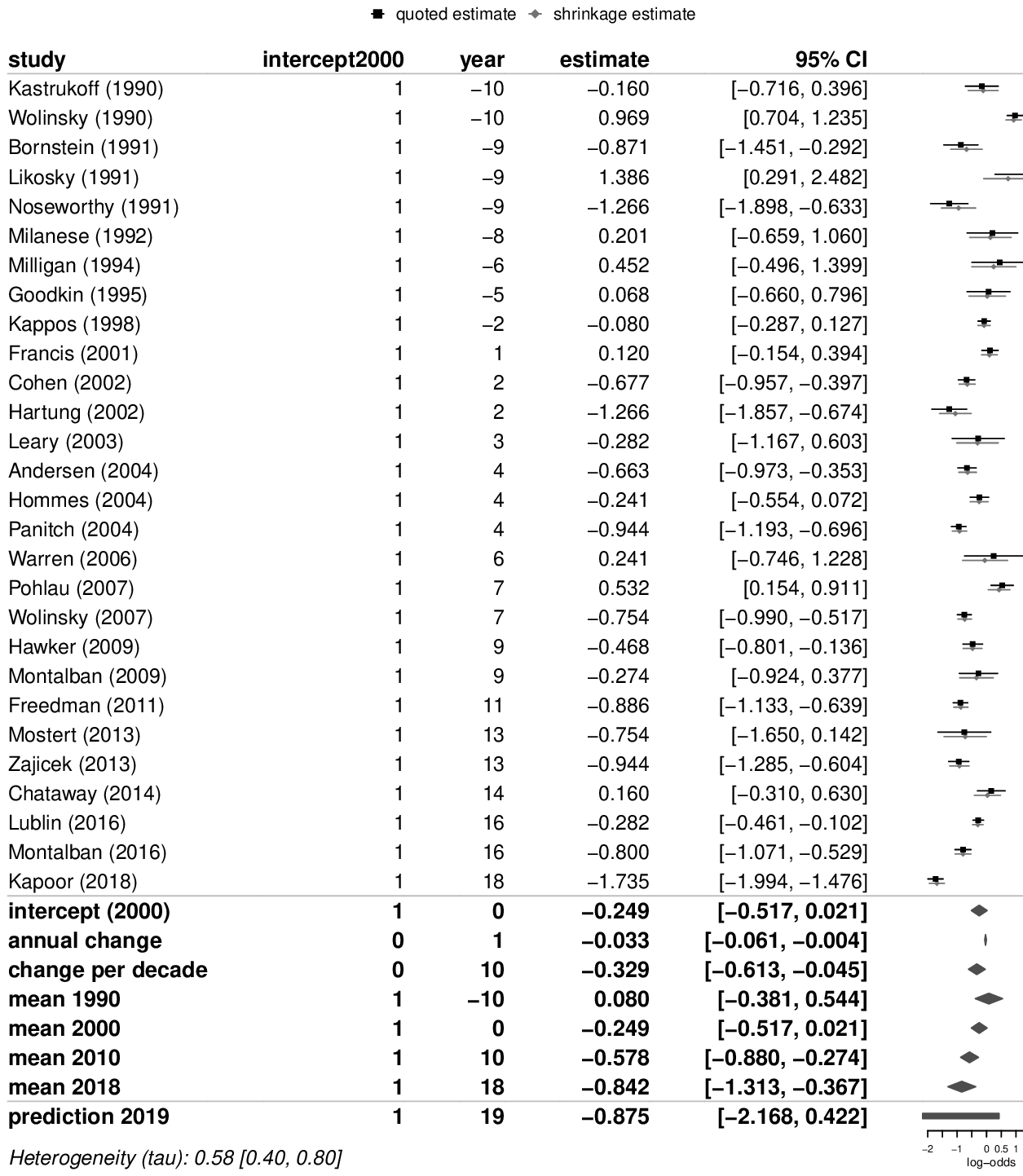}}
        \caption{\label{fig:NicholasForest} A forest plot illustrating the meta-regression results based on the multiple sclerosis data. The time trend is parameterized in terms of an \emph{intercept} and a \emph{year} variable that is centered at the year~2000. In addition to the ``plain'' intercept and (annual) slope, several linear combinations as well as a prediction for the year~2019 are also shown.}
      \end{figure}

      We may inspect the regression results based on the returned parameter estimates, but it may in fact be more illustrative to present these in a forest plot. Besides the two ``plain'' parameter estimates (intercept~$\beta_1$ and slope~$\beta_2$) we may check estimates of certain linear combinations corresponding to the mean at certain time points, or also to predicted values. A call of
\begin{lstlisting}[language=R]
R> forestplot(bmr07, xlab="log-odds",
+    X.mean=rbind("intercept (2000)"  =c(1,  0),
+                 "annual change"     =c(0,  1),
+                 "change per decade" =c(0, 10),
+                 "mean 1990"         =c(1,-10),
+                 "mean 2000"         =c(1,  0),
+                 "mean 2010"         =c(1, 10),
+                 "mean 2018"         =c(1, 18)),
+    X.predict=rbind("prediction 2019"=c(1, 19)))
\end{lstlisting}
      will generate a forest plot including the intercept and slope (annual change), the change per decade, the mean log-odds at several time points, as well as a prediction for the year~2019; the resulting plot is shown in Figure~\ref{fig:NicholasForest}.
      The annual change is estimated to be negative, implying a reduction in the log-odds by $0.033$ per year, or by $0.33$ per decade. For the odds this means a reduction by 3.2\% per year, or by 28\% per decade.

\begin{figure}[t!]
        \centering
        {\includegraphics[width=0.95\linewidth]{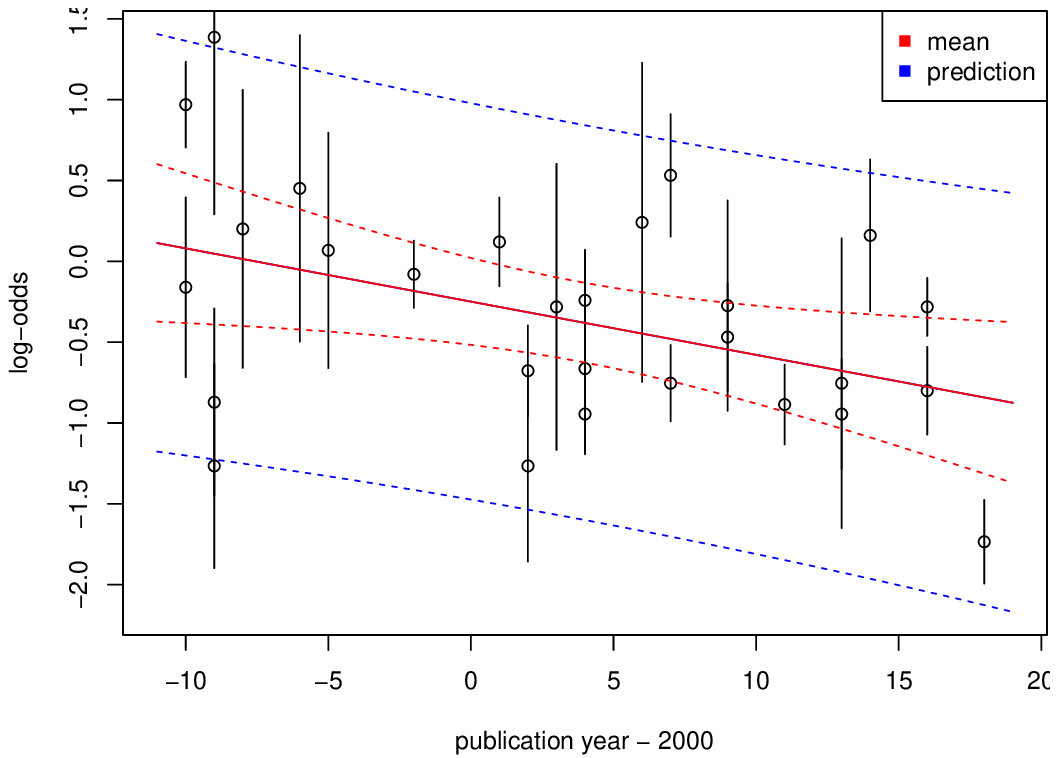}}
        \caption{\label{fig:NicholasTrend} A plot illustrating the multiple sclerosis data along with credible intervals for the mean as well as prediction intervals.}
      \end{figure}

      Besides the forest plot, it is often useful to illustrate the data along with the model estimates graphically. To that end, we can compute predictions and credible intervals and combine these in a single plot:
\begin{lstlisting}[language=R]
R> # specify corresponding "regressor matrix":
R> newx <- cbind(1, (1989:2019)-2000)
R> # compute credible intervals for the mean:
R> pred <- cbind("median"=bmr07$qpred(0.5,x=newx),
+                bmr07$pred.interval(x=newx))
R> # compute prediction intervals:
R> map <- cbind("median"=bmr07$qpred(0.5,x=newx,
+                                    mean=FALSE),
+               bmr07$pred.interval(x=newx,
+                                   mean=FALSE))
R> # show the 26 studies' point estimates
R> # and 95 percent CIs:
R> plot(es$year-2000, es$yi,
+       xlim=range(newx[,2]), ylim=range(map),
+       xlab="publication year - 2000",
+       ylab="log-odds")
R> matlines(rbind(es$year, es$year)-2000,
+           rbind(es$yi-qnorm(0.975)*sqrt(es$vi), 
+                es$yi+qnorm(0.975)*sqrt(es$vi)),
+           col=1, lty=1)
R> # show trend lines (and 95 percent intervals):
R> matlines(newx[,2],map,col="blue",lty=c(1,2,2))
R> matlines(newx[,2],pred,col="red",lty=c(1,2,2))
R> legend("topright",pch=15,col=c("red","blue"), 
+         c("mean","prediction"))
\end{lstlisting}
      The resulting trend plot is shown in Figure~\ref{fig:NicholasTrend}.
      All estimates (studies) along with their error bars are shown, and the estimated mean (along with credible and prediction intervals) is computed for a grid of values spanning the range from 1989 to 2019.

      \citet{NicholasEtAl2019} pointed out the good agreement with a subsequently published study by \citet{KapposEtAl2018} in~2018, who (depending on the exact definition used) reported between 23\% and 30\% progressing patients, corresponding to log-odds of~$-1.18$ or~$-0.83$, respectively.

      The predicted log-odds were -0.87 [-2.17, 0.42] for a ``future'' study in the year~2019 (see Figure~\ref{fig:NicholasForest}), corresponding to probabilities of 0.29 [0.10, 0.60].  Such a prediction could be useful for study design \citep{DerSimonian1996,FriedePohlmannSchmidli2019,GoudieEtAl2010} or sample size determination \citep{DeSantis2007}; it might also be utilized to supplement a study's sparse placebo data in terms of a meta-analytic-predictive (MAP) prior \citep{SchmidliEtAl2014}.


    \subsection{Several covariables}\label{sec:severalCovar}
      \citet{RobergeEtAl2017} performed a systematic literature review in order to summarize the evidence on effects of aspirin administered during pregnancy. Earlier research had already suggested that prophylactic administration of low-dose aspirin may reduce the prevalence of \emph{fetal growth restriction (FGR)}, which is a common cause of perinatal morbidity and mortality \citep{NardozzaEtAl2017}. While the exact mechanism by which aspirin works here is still unclear, it had become apparent that it is most effective when initiated early on, before 16~weeks of gestational age.

      A total of 35~studies were included in the eventual analysis; in 17~studies, therapy was initiated early ($\leq$16 weeks gestational age), and in 18~studies, onset was late ($>$16 weeks). Doses differed between studies and ranged from~$50$ up to~$150\,\mbox{mg}$ daily. For each study, we have numbers of cases and FGR events in treatment and control groups.

      We may load the example data and derive the log-ORs for all studies providing data on FGR events:
\begin{lstlisting}[language=R]
R> data("RobergeEtAl2017")
R> es.fgr <- escalc(measure="OR",
+    ai=asp.FGR.events,  n1i=asp.FGR.total,
+    ci=cont.FGR.events, n2i=cont.FGR.total,
+    slab=study, data=RobergeEtAl2017,
+    subset=
+      complete.cases(RobergeEtAl2017[,11:14]))
\end{lstlisting}
      At first we can check whether the aspirin dose appears to affect the chances of FGR; we can use the \texttt{model.matrix()} function to set up a corresponding regressor matrix and perform a simple analysis specifying an intercept and a linear effect for the dose. Again, due to the large number of studies included, we may utilize a non-informative (improper) uniform prior for the heterogeneity.
\begin{lstlisting}[language=R]
R> X01 <- model.matrix( ~ dose, data=es.fgr)
R> colnames(X01) <- c("intercept", "dose")
R> head(X01)
   intercept dose
01         1   50
02         1   60
03         1   60
04         1   60
06         1   75
07         1   75
> bmr08 <- bmr(es.fgr, X=X01)
> bmr08$summary
                 tau   intercept         dose
mode      0.27075394  0.08446175 -0.004604890
median    0.28349590  0.08799880 -0.004864835
mean      0.29005627  0.09099202 -0.005005075
sd        0.11591364  0.30196489  0.003685612
95% lower 0.06368288 -0.50870208 -0.012461472
95% upper 0.52626505  0.69503106  0.002157158
\end{lstlisting}
      So far, this does not convincingly indicate an effect; while the ``dose'' effect is estimated to be negative, implying a reduction in FGR events with increasing dose, the 95\% CI includes both negative as well as positive values.

      Considering the earlier suggestion of the relevance of the timepoint of therapy initiation, we may then check whether the effect might differ between studies implementing an ``early'' or ``late'' onset. Such a model may be defined in different ways; here we will consider a setup including individual intercepts and slopes in both groups of studies:
\begin{lstlisting}[language=R]
R> X02 <- model.matrix(~ -1 + onset + onset:dose,
+                      data=es.fgr)
R> colnames(X02) <- c("intEarly", "intLate", 
+                     "doseEarly", "doseLate")
R> head(X02)
   intEarly intLate doseEarly doseLate
01        1       0        50        0
02        1       0        60        0
03        1       0        60        0
04        1       0        60        0
06        1       0        75        0
07        1       0        75        0
\end{lstlisting}
      We can see that the first studies within the data set all belong to the ``early'' group; in the second group, the regressor matrix entries 
      corresponding
      to the ``late'' effects then are non-zero instead. We may then run the analysis based on the extended model:
\begin{lstlisting}[language=R]
R> bmr09 <- bmr(es.fgr, X=X02)
R> bmr09$summary[,1:3]
                 tau   intEarly     intLate
mode      0.08758667  0.4290267  0.05426766
median    0.11720808  0.4263765  0.04654656
mean      0.13062301  0.4248359  0.04167659
sd        0.09016564  0.5201030  0.24672617
95% lower 0.00000000 -0.5970099 -0.45148496
95% upper 0.29774638  1.4449251  0.52534390
R> bmr09$summary[,4:5]
             doseEarly     doseLate
mode      -0.014014415 -0.001538704
median    -0.013981904 -0.001491126
mean      -0.013970065 -0.001474899
sd         0.006281788  0.002990934
95% lower -0.026280559 -0.007383067
95% upper -0.001626998  0.004450449
\end{lstlisting}
      From the analysis results, we now see a somewhat different picture; first of all, the heterogeneity ($\tau$) is reduced, from a median of 0.28 down to 0.12. The ``late'' slope parameter still is small and centered near zero, while the ``early'' slope parameter along with its 95\% CI is on the negative side.

\begin{figure}[b!]
        \centering
        {\includegraphics[width=0.95\linewidth]{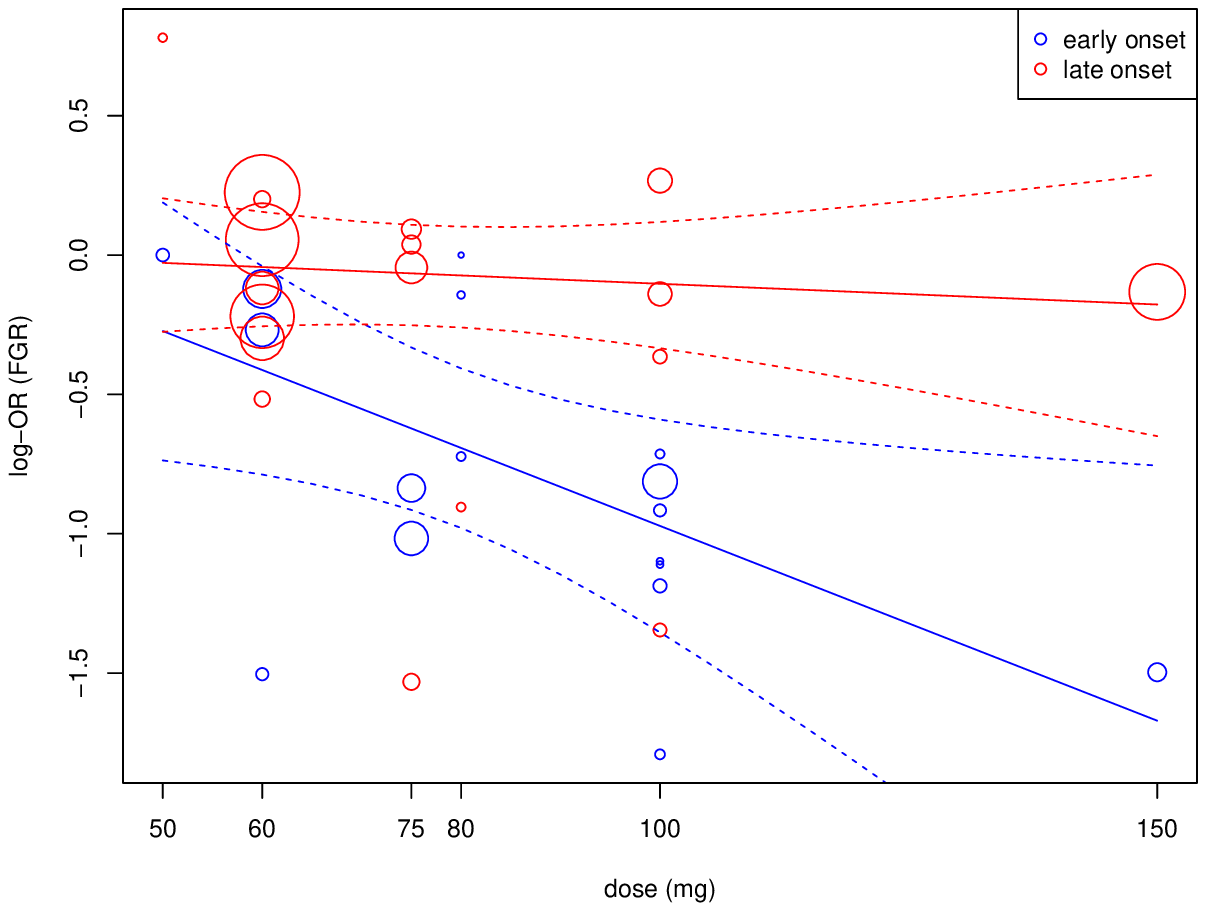}}
        \caption{\label{fig:RobergeTrend} A ``bubble plot'' illustrating the data along with credible intervals for the mean effect as functions of the dose in ``early'' and ``late'' study groups. The point sizes are inversely proportional to the standard errors.}
      \end{figure}

      We may illustrate data and regression lines jointly in a ``bubble plot''; the estimated ORs as functions of the regressors may again be extracted from the \texttt{bmr()} function's output:
\begin{lstlisting}[language=R]
R> # derive predictions from the model;
R> # specify corresponding "regressor matrices":
R> newx.early <- cbind(1, 0, seq(50,150,by=5), 0)
R> newx.late  <- cbind(0, 1, 0, seq(50,150,by=5))
R> # compute predicted medians 
R> # and 95 percent intervals: 
R> pred.early <- cbind("median"=bmr09$qpred(0.5,
+                                  x=newx.early),
+                bmr09$pred.interval(x=newx.early))
R> pred.late <- cbind("median"=bmr09$qpred(0.5, 
+                                  x=newx.late),
+               bmr09$pred.interval(x=newx.late))
R> # draw "bubble plot": 
R> plot(es.fgr$dose, es.fgr$yi, 
+    cex=1/sqrt(es.fgr$vi),
+    col=c("blue","red")[as.numeric(es.fgr$onset)],
+    xlab="dose (mg)", ylab="log-OR (FGR)")
R> legend("topright", col=c("blue","red"), 
+    c("early onset", "late onset"), pch=1)
R> # add predictions to bubble plot:
R> matlines(newx.early[,3], pred.early, 
+    col="blue", lty=c(1,2,2))
R> matlines(newx.late[,4], pred.late,
+    col="red", lty=c(1,2,2))
\end{lstlisting}
      The resulting trend plot is shown in Figure~\ref{fig:RobergeTrend}.
      ``Larger'' studies (those with smaller standard errors) are denoted by larger symbols.  
      
      Again, a range of alternative model or prior specifications may be sensible here;
      e.g., different parametrizations (common intercept or slope parameters, individual heterogeneity parameters, differing prior specifications, a zero intercept, \ldots), which may also suggest the use of a \emph{model selection} approach (see following Section~\ref{sec:modelselection}).  In addition, it may also be interesting to investigate the sensitivity of results to individual data points (studies).


    \subsection{Model selection}\label{sec:modelselection}
      \subsubsection{The example setup}
        \citet{CinarEtAl2021} discussed a meta-analysis problem involving a total of four potential covariables. Their example data set included 80~studies investigating biomass production of maize plants under different conditions; of interest were the effects of inoculation using symbiotic mycorrhizal fungi. Four dichotomous aspects varied between the studies, namely the type of fungus (FUN, \emph{funneliformis} or \emph{rhizophagus}), the use of phosphorus fertilizer (FP, yes/no), use of nitrogen fertilizer (FN, yes/no), and sterilization of the soil (STER, yes/no). The endpoint was expressed in terms of a logarithmic response ratio \citep{HedgesGurevitchCurtis1999}, and the problem was first of all to determine \emph{which} of the variables affected the yield, making this a \emph{variable selection} or \emph{model selection} problem.

        In the present case, combinations of the four (binary) variables allow to specify 16~different \emph{models}, ranging from the model without covariables (besides an overall intercept) to the model with all four included. In a Bayesian context, after specification of prior probabilities (for all models themselves, as well as for parameters within models), one may then derive posterior probabilities for each model, or one may compare and rank models based on their associated \emph{Bayes factors} \citep{BergerPericchi2001,BDA3rd,KassRaftery1995,SpiegelhalterEtAl}. The uncertainty involved in the model selection may also be accounted for (or in fact, the \emph{selection} of a single model is avoided) by using a \emph{model averaging} approach \citep{Clemen1989,ClydeGeorge2004,HoetingEtAl1999,RafteryMadiganHoeting1997,RoeverWandelFriede2018}. Another approach may be to consider the \emph{median probability model} based on all variables' marginal inclusion probabilities \citep{BarbieriBerger2004,BarbieriBergerEtAl2021}. Either way, computations hinge on the determination of \emph{marginal likelihoods}, which first of all is often computationally challenging, and secondly, requires the specification of \emph{proper} priors for all parameters within the 16~models. Unlike in many parameter estimation problems, the exact details of (non-informative or weakly informative) prior specifications are crucial and may affect results in sometimes unintuitive ways, as exemplified in \emph{Lindley's paradox} \citep{BDA3rd,Lindley1957}, so that particular caution is advised here.  In some cases, it may be worth considering whether a model selection approach is in fact the method of choice \citep{GelmanRubin1995}.


      \subsubsection{Model specification}
        We assume all of the 16~possible models to be a~priori equally likely; within each model we then assign a vague prior for the intercept (normal with mean zero and standard deviation~10), and weakly informative priors for the binary covariables' effects (normal with mean zero and standard deviation~2.82). The effect prior confines the likely effect magnitudes on their logarithmic scale so that back on the (exponentiated) scale of response ratios these roughly correspond to values within factors of~$250$ and~$\frac{1}{250}$ \citep{GunhanRoeverFriede2020}. For the heterogeneity parameter ($\tau$), we assume a weakly informative half-normal distribution with scale~$0.5$ \citep{RoeverEtAl2021}.

        Note that equal prior probabilities for all models imply that all variables a-priori have 50\% inclusion probability, and that the prior expected number of parameters is~$\frac{N}{2}$ (where $N$~is the total number of variables).  Alternatively, different specifications are also conceivable; for example, assigning a probability~$\pi$ for each variable to be included implies a probability $\pi^n (1-\pi)^{(N-n)}$ for each single model (where $n$ is the number of variables included) and it implies a~priori a binomially distributed total number of included parameters (with expectation $\pi N$).


      \subsubsection{Implementation}
        The example data are available in \citeauthor{CinarEtAlData2020}'s online supplement \citep{CinarEtAlData2020}. We may download the data, read them into~\textsf{R}, and then systematically apply the 16~possible meta-regression models. Computations may take a few minutes.
\begin{lstlisting}[language=R]
R> # load data:
R> CinarEtAl2021 <- read.csv("CinarEtAl2021.csv")
R> # convert effect sizes:
R> effsize <- escalc(measure="ROM", yi=yi, vi=vi, 
+                    data=CinarEtAl2021)
R> head(effsize)
  FUN FP FN STER      yi     vi 
1   0  0  0    0  0.0950 0.4000 
2   0  0  0    0 -1.8590 0.4000 
3   0  0  0    0  0.3640 0.6667 
4   0  0  0    0  0.2880 0.4000 
5   0  0  0    0  0.1310 0.2500 
6   0  0  0    0  0.0000 0.4000 
R> # figure out possible models
R> # (factor combinations):
R> models <- expand.grid("FUN"=c(FALSE,TRUE), 
+                        "FP"=c(FALSE,TRUE),
+                        "FN"=c(FALSE,TRUE),
+                        "STER"=c(FALSE,TRUE))
R> models <- as.matrix(models)
R> rownames(models) <- LETTERS[1:16]
R> # generate list of 16 regressor matrices:
R> Xlist <- vector(nrow(models), mode="list")
R> names(Xlist) <- rownames(models)
R> for (i in 1:nrow(models)) {
+    Xlist[[i]] <- matrix(
+      rep(1, nrow(CinarEtAl2021)),
+      ncol=1, dimnames=list(NULL, "intercept"))
+    if (any(models[i,])) {
+      Xlist[[i]] <- cbind(Xlist[[i]], 
+    as.matrix(CinarEtAl2021[,1:4])[,models[i,]])
+      colnames(Xlist[[i]]) <- c("intercept", 
+        colnames(models)[models[i,]])
+    }
+  }
R> # perform 16 regression analyses:
R> bmrlist <- vector(nrow(models), mode="list")
R> names(bmrlist) <- rownames(models)
R> for (i in 1:nrow(models)) {
+    bmrlist[[i]] <- bmr(effsize, X=Xlist[[i]],
+      tau.prior=function(t){dhalfnormal(t,0.5)},
+      beta.prior.mean=rep(0, ncol(Xlist[[i]])),
+      beta.prior.sd=c(10,
+                      rep(2.82,
+                          ncol(Xlist[[i]])-1)))
+  }
\end{lstlisting}
        The 16~regression outputs are now stored in the ``\texttt{bmrlist}'' object. The marginal likelihood for a \texttt{bmr()} output is stored in the ``\texttt{\ldots\$marginal.likelihood}'' element (presuming that proper priors have been used for the analysis). We may now assemble these numbers, and combine them with the models' prior probabilities to derive the posterior probabilities.
\begin{lstlisting}[language=R]
R> # specify model prior probabilities:
R> priorprob <- rep(1/16, 16)
R> # assemble marginal likelihoods
R> margi <- sapply(bmrlist,
+    function(x){x$marginal.likelihood})
R> # determine posterior model probabilities: 
R> prob <- (priorprob*margi)/sum(priorprob*margi)
\end{lstlisting}

        Table~\ref{tab:CinarProbs} illustrates the 16~models along with their posterior probabilities. The a-posteriori most probable model at the top of the list is the one including (besides an intercept) the FP and FN variables, corresponding to influential effects for phosphorus and nitrogen fertilizers.

        \begin{table}[b!]
        \begin{center}
          \caption{\label{tab:CinarProbs} The 16~models and their probabilities (in descending order). A dot (\Vin) indicates that a variable is included in a model, an open circle (\Vout) means that it is not included. The very last line shows the four variables' marginal inclusion probabilities.} \vspace{1ex}
          \begin{tabular}{rccccc} 
            \toprule
            & \multicolumn{4}{c}{included variables} & \\
            \cmidrule(lr){2-5}
            model & FUN & FP & FN & STER & probability \\
            \midrule
             1  & \Vout & \Vin  & \Vin  & \Vout & 0.6293 \\
             2  & \Vout & \Vin  & \Vin  & \Vin  & 0.1076 \\
             3  & \Vin  & \Vin  & \Vin  & \Vout & 0.0907 \\
             4  & \Vin  & \Vin  & \Vout & \Vout & 0.0645 \\[1ex]
             5  & \Vout & \Vin  & \Vout & \Vin  & 0.0380 \\
             6  & \Vin  & \Vin  & \Vout & \Vin  & 0.0304 \\
             7  & \Vin  & \Vin  & \Vin  & \Vin  & 0.0148 \\
             8  & \Vout & \Vin  & \Vout & \Vout & 0.0106 \\[1ex]
             9  & \Vin  & \Vout & \Vout & \Vout & 0.0077 \\
            10  & \Vin  & \Vout & \Vout & \Vin  & 0.0019 \\
            11  & \Vout & \Vout & \Vout & \Vin  & 0.0014 \\
            12  & \Vout & \Vout & \Vout & \Vout & 0.0011 \\[1ex]
            13  & \Vin  & \Vout & \Vin  & \Vout & 0.0009 \\
            14  & \Vout & \Vout & \Vin  & \Vout & 0.0007 \\
            15  & \Vin  & \Vout & \Vin  & \Vin  & 0.0002 \\
            16  & \Vout & \Vout & \Vin  & \Vin  & 0.0002 \\
            \midrule
                & 0.2111 & 0.9859 & 0.8444 & 0.1946 & \\ 
            \bottomrule
          \end{tabular}
        \end{center}
        \end{table}

        The \emph{median probability model}, i.e., the model including all those variables that have a marginal inclusion probability $\geq$0.5 \citep{BarbieriBerger2004,BarbieriBergerEtAl2021}, here also coincides with the most probable model. The three most probable models also match the top~3 models based on the \emph{Akaike information criterion (AIC)} as quoted by \citet{CinarEtAl2021}.

        Table~\ref{tab:CinarEsti} shows the parameter estimates from the most probable model.
        Instead of singling out one ``best'' model for inference, one might now also utilize the results in a \emph{model averaging} approach, effectively using all 16~models simultaneously and weighting predictions based to their associated probabilities \citep{Clemen1989,ClydeGeorge2004,HoetingEtAl1999,RafteryMadiganHoeting1997,RoeverWandelFriede2018}.

        It should be noted that, in a sense, the example shown here was particularly ``simple'' since all variables considered were in the same ``units'' (binary), so that it is relatively easy to specify a neutral prior without favouring any of the variables from the start. 
        \begin{table}[t]
        \begin{center}
          \caption{\label{tab:CinarEsti}Parameter estimates for the most probable model including the FP and FN variables (phosphorus and nitrogen fertilizers), which receives a posterior probability of~0.63.} \vspace{1ex}
          \begin{tabular}{lcc} 
            \toprule
            parameter & median & 95\% CI \\
            \midrule
            heterogeneity ($\tau$) & 0.510 & [0.339, 0.690] \\
            intercept ($\beta_1$) & 0.227 & [-0.097, 0.652] \\
            FP ($\beta_2$) & -1.006 & [-1.416, -0.582] \\
            FN ($\beta_3$) & 0.894 & [0.429, 1.346] \\
            \bottomrule
          \end{tabular}
        \end{center}
        \end{table}        
        In practice, it might also be of interest to check the results' sensitivity to any of the prior specifications, or to also investigate the possible relevance of interaction effects (as a simple additive effect of the two fertilizers may or may not be biologically plausible).

        In the model selection context, the use of \emph{penalized complexity priors} \citep{KleinKneib2016} may also play a more prominent role than in ``simple'' meta-analysis applications. Penalized complexity priors here correspond to exponential priors for the heterogeneity~($\tau$) \citep{RoeverEtAl2021}.


  \section{Discussion} \label{sec:discussion}
    In the present article, we demonstrated the use of Bayesian meta-regression as facilitated through the \texttt{bayesmeta} \textsf{R}~package \citep{bayesmeta}. The implementation is conveniently based on the \textsc{direct} algorithm \citep{RoeverFriede2017} and constitutes a straightforward generalisation of ``simple'' meta-analysis within the NNHM framework \citep{Roever2020}. This way, a wide range of  extensions such as subgroup analysis, continuous covariables, indirect comparisons, or model selection are covered.
    
    While Bayesian analyses sometimes tend to be technically demanding, through to its user-friendly interface, its generality and the quick and reproducible computation, the \texttt{bayesmeta} implementation provides a low-threshold entry point for a wider audience beyond computational experts. 
    While a certain amount of preparation certainly is still required, it is relatively easy to extend an existing \texttt{bayesmeta} implementation to include covariables in addition, whereas the effort required to run, diagnose and possibly implent an MCMC approach would be substantially higher.
    In many applications, use of the \texttt{bmr()}, \texttt{forestplot()} and \texttt{summary()} functions may already be sufficient to address most relevant questions. More sophisticated investigations are possible using the comprehensive output available (such as custom plots (see Sections~\ref{sec:exampleNetworkMA}--\ref{sec:severalCovar}), model selection (see Section~\ref{sec:modelselection}) or model averaging \citep{RoeverWandelFriede2018}).
    
    The meta-regression approach presented here builds on the NNHM, which yields accurate inference in particular also in case of few studies \citep{FriedeRoeverWandelNeuenschwander2017a}. A normal approximation at the study-level is often appropriate, for example when sample sizes are ``large'' and and studies are sufficiently powered. However, the normal approximation may also deteriorate in certain circumstances, for example, for binary endpoints with zero (or near-zero) event counts. 
    In such cases, use of an exact likelihood may be preferable \citep{StijnenEtAl2010,Tu2014}, as for example implemented in the \texttt{metaStan} \textsf{R}~package \citep{GunhanRoeverFriede2020}. This package also allows for \emph{model-based meta-analysis (MBMA)}, where studies may contribute information on several (more than two) study arms that correspond to different exposures or dose levels \citep{GuenhanRoeverFriede2022,MawdsleyEtAl2016}. 
    Similarly, the \texttt{bmeta} package provides meta-analysis and meta-regression functionalities based on MCMC sampling \citep{R:bmeta}.
    The \texttt{RBesT} package also supports meta-analysis based on a range of endpoint types, with a focus on deriving (``meta-analytic-predictive (MAP)'') prior distributions for use in a subsequent analysis \citep{WeberEtAl2021}. 
    In the example applications we included indirect comparisons as a very basic example of a treatment network; more complex models commonly applied for network meta-analysis (see e.g. \citep[Sec.~11]{CochraneHandbook}) are currently not implemented in the \texttt{bayesmeta} package. Dedicated packages are available for network meta-analyis, for instance \texttt{nmaINLA}, which utilizes the \emph{integrated nested Laplace approximation (INLA)} for posterior inference \citep{GuenhanEtAl2018}.
    Recently, \citet{WilliamsRodriguezBuerkner2021} proposed meta-anaytic models with covariate effects on the heterogeneity variance besides the mean; these are implemented in the \textsf{R}~package \texttt{blsmeta}.
    The \texttt{bspmma} and \texttt{metaBMA} \textsf{R}~packages implement extensions of the ``simple'' NNHM, but currently without the option for meta-regression \citep{R:bspmma,R:metaBMA}. 
    Alternatively, many meta-analysis problems may also be formulated in terms of generalized linear mixed models (GLMMs), which can be fitted for instance using the \textsf{R}~package \texttt{brms} \citep{Buerkner2017}.
    Most of the above alternative Bayesian packages rely on MCMC methods for inference. In case a taylored solution is required, which is not covered by any of the mentioned packages, it is probably easiest to also resort to MCMC methods, which may be implemented e.g.\ using the JAGS or Stan engines, and the interfacing \texttt{rjags} or \texttt{rstan} packages \citep{Plummer2003,R:rjags,R:rstan}.
    For a comprehensive and up-to-date overview of available \textsf{R}~packages, see also the corresponding \emph{CRAN task view} \citep{CranTaskViewMA}.


  \section*{Acknowledgment}
  Support from the \emph{Deutsche For\-schungs\-ge\-mein\-schaft (DFG)} is
  gratefully acknowledged (grant number \mbox{FR~3070/3-1}).

  \section*{Conflict of interest statement}
  The authors declare no conflict of interest.

  \section*{Ethics approval}
  Not applicable.


  \bibliographystyle{abbrvnat}
  \bibliography{literature}

\end{document}